\documentclass[aps,prd,superscriptaddress]{revtex4-2}

\usepackage{enumerate}

\usepackage{amsmath}
\usepackage{amssymb}
\usepackage{amsfonts}
\usepackage{amscd} % 画交换图用
\usepackage{mathrsfs} % 提供了mathscr 命令

\usepackage{latexsym}
\usepackage{bm}

\usepackage{graphicx}
\usepackage{epsfig}
\usepackage{hhline,multirow}
\usepackage{dcolumn}
\usepackage{url}
\usepackage[colorlinks=true,linkcolor=blue]{hyperref}
\usepackage{color}
\usepackage{indentfirst}
\usepackage{subfigure}
\usepackage{psfrag}
\usepackage{slashed}
\usepackage{float}
\usepackage{setspace}
\usepackage{multirow}

\newcommand{\mev}{\,\mathrm{MeV}}

\DeclareMathOperator{\tr}{Tr} 
\newcommand*{\dif}{\mathop{}\!\mathrm{d}}

\begin{document}

\title{Electromagnetic form factors of octet baryons with the nonlocal chiral effective theory}
\author{Mingyang Yang}
\affiliation{Institute of High Energy Physics, CAS, P. O. Box
	918(4), Beijing 100049, China}
\affiliation{School of Physical Sciences, University of Chinese Academy of Sciences, Beijing 101408, China}

\author{P. Wang}
\affiliation{Institute of High Energy Physics, CAS, P. O. Box
	918(4), Beijing 100049, China}
\affiliation{Theoretical Physics Center for Science Facilities,
	CAS, Beijing 100049, China}

\begin{abstract}

The electromagnetic form factors of octet baryons are investigated with the nonlocal chiral effective theory.  The nonlocal interaction generates both the regulator which makes the loop integral convergent and the $Q^2$ dependence of form factors at tree level.  Both octet and decuplet intermediate states are included in the one loop calculation. The momentum dependence of baryon form factors is studied up to 1 GeV$^2$ with the same number of parameters as for the nucleon form factors. The obtained magnetic moments of all the octet baryons as well as the radii are in good agreement with the experimental data and/or lattice simulation.

\end{abstract}
%%% \pacs{13.40.Gp; 13.40.Em; 12.39.Fe; 14.20.Dh}
\maketitle
\section{Introduction}

The study of electromagnetic form factors of hadrons is of crucial importance to understand their sub-structure. A lot of theoretical and experimental efforts have been made in this field. On the one hand, with the upgrade of experimental facilities, the parton distribution functions (PDFs) from the deep inelastic scattering as well as the form factors at relatively large momentum transfer from the elastic scattering can be extracted \cite{Camsonne,CMS}. On the other hand, many measurements on form factors have been carried out at very small momentum transfer to get the information of the nucleon radii as accurate as possible \cite{Xiong,Bernauer}.

Theoretically, though QCD is the fundamental theory to describe strong interactions, it is difficult to study hadron physics using QCD directly. There are many phenomenological models, such as the cloudy bag model \cite{theory.bag-model}, the constituent quark model \cite{theory.Chiral-Constituent-Quark-Model}, the 1/Nc expansion approach \cite{theory.1diNc}, the Nambu-Jona-Lasino (NJL) model \cite{Ito}, the perturbative chiral quark model (PCQM) \cite{theory.chiral-quark.model}, the extended vector meson dominance model \cite{theory.extended-vector-meson}, the SU(3) chiral quark model \cite{theory.extended-chiral-quark-model}, the quark-diquark model \cite{theory.quark-diquark}, etc. These model calculations are helpful to provide the physical scenario for the hadron structure.

Besides the phenomenological quark models, there are two systematic methods in hadron physics. One is the lattice simulation and the other is an effective field theory (EFT) of QCD, chiral perturbation theory (ChPT).  Historically, most formulations of ChPT are based on dimensional or infrared regularization (IR). Though ChPT is a successful approach, for the nucleon electromagnetic form factors, it is only valid for $Q^2 < 0.1$ GeV$^2$ \cite{theory.chpt-range.1}. 
When vector mesons are included, the result is close to the experiments when $Q^2$ is less than 0.4 GeV$^2$ \cite{theory.chpt-range.2}. An alternative regularization method, namely finite-range-regularization (FRR) has been proposed. Inspired by quark models that account for the finite-size of the nucleon as the source of the pion cloud, effective field theory with FRR has been widely applied to extrapolate lattice data of  vector meson mass, magnetic moments, magnetic form factors, strange form factors, charge radii, first moments of GPDs, nucleon spin, etc \cite{theory.FRR.0,theory.FRR.1,theory.FRR.2,theory.FRR.3,theory.FRR.4,theory.FRR.5,theory.FRR.6,theory.FRR.7,theory.FRR.8,Wang}. 

Recently, we proposed a nonlocal chiral effective Lagrangian which makes it possible to study the hadron properties at relatively large $Q^2$ \cite{fw.nucleon,fw.strange}. The nonlocal interaction generates both the regulator which makes the loop integral convergent and the $Q^2$ dependence of form factors at tree level. The obtained electromagnetic form factors and strange form factors of the nucleon are very close to the experimental data \cite{fw.nucleon,fw.strange}. This nonlocal chiral effective theory was also applied to study the parton distribution functions and Sivers functions of the sea quarks in nucleons \cite{Salamu.parton,He}. In addition, the nonlocal behavior is further assumed to be a general property for all the interactions and an example of this assumption is the application to the lepton anomalous magnetic moments \cite{He2}. 

Since the nonlocal effective theory provides good descriptions of the nucleon form factors up to relatively large momentum transfer, in this paper, we will extend our study from nucleon to all the octet baryons. While the nucleon form factors are precisely determined experimentally, those of the other octet baryons are significantly more challenging to measure and as a result are poorly known from nature. Compared with the experiments, for the lattice gauge theory, it is not very difficult to extend the simulation of the nucleon form factors to the other octet form factors. Some lattice simulations on octet form factors have been reported \cite{Boinepalli,Shanahan,Shanahan2}. 

The form factors of octet baryons were also studied in heavy baryon and relativistic chiral perturbation theory with different regularization schemes. In Ref.~\cite{Kubis}, the magnetic moments and electromagnetic radii of octet baryons were calculated in relativistic ChPT with infrared regularization. The electromagnetic form factors up to 0.4 GeV$^2$ were further studied with the inclusion of vector mesons. The magnetic moments of octet baryons were also studied in relativistic ChPT with extendend-on-mass-shell (EOMS) renormalization scheme, where the intermediate decuplet states were not included \cite{Xiao}. In Refs.~\cite{Geng,Blin}, the decuplet states were included in the calculation of octet-baryon form factors with EOMS scheme. In particular, vector mesons were included explicitly to improve the final results in Ref.~\cite{Blin}.

Here, we will apply the nonlocal chiral effective theory to investigate the electromagnetic form factors of all the octet baryons up to 1 GeV$^2$ as well as the magnetic moments and radii. The paper is organized as follows. In section II, we will introduce the nonlocal chiral Lagrangian and the expressions for the form factors are written in appendix. Numerical results are presented in section III and finally, section IV is a short summary.

\section{Formalism}
	
The lowest order chiral Lagrangian for baryons, pseudo-scalar mesons and their interactions can be written as \cite{fw.nucleon,Salamu.parton,Jenkins1,Jenkins2}
	\begin{equation}\label{eq:chiral-lagrangian}
		\begin{aligned}
			\mathcal{L} &{}=
			i \tr\left(\bar{B} \gamma_{\mu} \slashed{\mathscr{D}}^\mu B\right) -m_B \tr\left(\bar{B}B\right)
			{}+{}\bar{T}_\mu^{abc}(i\gamma^{\mu\nu\alpha}D_\alpha-m_T\gamma^{\mu\nu})
			T_\nu^{abc}
			+\frac{f^2}{4}\tr\left(\partial_\mu U \partial^\mu U^\dagger\right)
			\\ &{}
			+D\tr\left(\bar{B} \gamma_\mu \gamma_5 \{A^\mu,B\}\right)
			+ F\tr\left(\bar{B} \gamma_\mu \gamma_5 [A^\mu,B]\right)
			%%%\\ &{}
			+\frac{\mathcal{C}}{f}{\epsilon}^{abc}
			\bar{T}_{\mu,a}^{\phantom{\mu,a}de}
			(g^{\mu\nu}+ z \gamma^\mu\gamma^\nu) B_{ce}\partial_\nu\phi_{bd}+H.C,
		\end{aligned}
	\end{equation}
	where $D$, $F$ and $ {\cal C} $ are the coupling constants. $z$ is the off-shell parameter.
	The chiral covariant derivative $\mathscr{D}_\mu$ is defined as $\mathscr{D}_\mu
	B=\partial_\mu B+[V_\mu,B]$. The pseudo-scalar meson octet
	couples to the baryon field via the vector and axial-vector
	combinations as
	\begin{equation}
		\begin{aligned}
			V_\mu=\frac{1}{2}(\zeta\partial_\mu\zeta^\dagger+\zeta^\dagger\partial_\mu\zeta)+\frac{1}{2}ie\mathscr{A}_\mu
			(\zeta^\dagger Q_c \zeta+\zeta Q_c \zeta^\dagger), \quad
			A_\mu=\frac{i}{2}(\zeta\partial_\mu\zeta^\dagger-\zeta^\dagger\partial_\mu\zeta)-\frac{1}{2}e\mathscr{A}_\mu
			(\zeta Q_c \zeta^\dagger-\zeta^\dagger Q_c \zeta),
		\end{aligned}
	\end{equation}
	where
	\begin{equation}
		\zeta^2 = U =e^{i2\phi/f},\qquad f=93~\mev.
	\end{equation}
	$Q_c$ is the real charge matrix $\text{diag} (2/3,-1/3,-1/3)$.
	$\phi$ and $B$ are the matrices of pseudo-scalar fields and octet baryons.
	$\mathscr{A}_\mu$ is the photon field.
	The covariant derivative $D_\mu$ in the decuplet sector is defined as
	$D_\nu T_\mu^{abc} = \partial_\nu T_\mu^{abc}+(\Gamma_\nu,T_\mu)^{abc}$,  
	where $\Gamma_\nu$ 
	is the chiral connection defined as
	$(X,T_\mu)^{abc}=(X)_d^aT_\mu^{dbc}+(X)_d^bT_\mu^{adc}+(X)_d^cT_{\mu}^{abd}$. 
	$\gamma^{\mu\nu\alpha}$, $\gamma^{\mu\nu}$ are the antisymmetric matrices expressed as
	\begin{equation}
		\gamma^{\mu\nu}
		=\frac12\left[\gamma^\mu,\gamma^\nu\right]\hspace{.5cm}\text{and}\hspace{.5cm}
		\gamma^{\mu\nu\rho}=\frac14\left\{\left[\gamma^\mu,\gamma^\nu\right],
		\gamma^\rho\right\}.
	\end{equation}
	
The octet, decuplet and octet-decuplet transition operators for magnetic moment are needed in the one loop calculations. 
The anomalous magnetic Lagrangian of octet baryons is written as
	\begin{equation}
		\label{eq.octet.anomalous}
		\begin{aligned}
			\mathcal{L}_{\text{oct}}=\frac{e}{4m_B}
			\Big(
			c_1 \tr\left(\bar{B} \sigma^{\mu\nu}
			\left\{F^+_{\mu\nu},B\right\}\right) +
			c_2 \tr\left(\bar{B}\sigma^{\mu\nu} \left[F^+_{\mu\nu},B\right]\right)+
			c_3 \tr\left(\bar{B}\sigma^{\mu\nu}B\right)\tr\left(F^+_{\mu\nu}\right)
			\Big),
		\end{aligned}
	\end{equation}
	where
	\begin{equation}
		F^\dagger_{\mu\nu}=-\frac12\left(\zeta^\dag F_{\mu\nu} Q_c \zeta+\zeta
		F_{\mu\nu} Q_c \zeta^\dag\right).
	\end{equation}
	The above Lagrangian will contribute to the Pauli form factor $F_2$ which is defined in Eq.~(\ref{eq:f1f2}).
	At the lowest order, the contribution of quark $q$ with unit charge to the 
	octet magnetic moments can be obtained by replacing the 
	charge matrix $Q_c$ with the corresponding diagonal quark 
	matrices $\lambda_q = \text{diag}(\delta_{qu}, \delta_{qd}, \delta_{qs})$.
	Let's take the nucleon as an example.
	After expansion of the above equation, it is found that
	\begin{equation}\label{treemag.chpt}
		\begin{aligned}
		&F_2^{p,u}=c_1+c_2+c_3,&\quad
		&F_2^{p,d}=c_3,&\quad
		&F_2^{p,s}=c_1-c_2+c_3,&\quad \\
		%%%+++++++++++++++++++++++++++++++++++++++++++++
		&F_2^{n,u}=c_3,&\quad
		&F_2^{n,d}=c_1+c_2+c_3,&\quad
	   	&F_2^{n,s}=c_1-c_2+c_3.&\quad
		\end{aligned}
	\end{equation}
	Comparing with the results of the constituent quark model where
	\begin{equation}
		\label{treemag.quarkmodel}
		\begin{aligned}
			F_2^{p,s}={}& 0 ~~\text {and}~~
			F_2^{n,s}=0, \\
		\end{aligned}
	\end{equation}
	we get
	\begin{equation}
		c_3=c_2-c_1.
	\end{equation}
	The decuplet anomalous magnetic moment operator is expressed as
	\begin{equation}\label{eq:ci}
		\mathcal{L}_{\text{dec}}=
		-\frac{ieF_2^T}{4M_T}
		\bar{T}_{\mu,abc}\sigma_{\rho\lambda}F^{\rho\lambda}T^{\mu,abc}.
	\end{equation}
	The transition magnetic operator is written as
	\begin{equation}
		\mathcal{L}_{\text{trans}}=
		i\frac{e}{4m_B}\mu_T F_{\mu\nu}
		\Big(
		\epsilon^{ijk} Q_{c,il}\bar{B}_{jm}\gamma^\mu\gamma_5
		T^{\nu,klm} 
		+%%%%%%%%%%%%%%%%%%%%%%%%%%%
		\epsilon^{ijk} Q_{c,li}\bar{T}^{\mu,klm} \gamma^\nu\gamma_5 B_{mj}
		\Big).
	\end{equation}
	The anomalous magnetic moments of baryons can also also be expressed in terms of quark magnetic moments $\mu_q$.
	For example, $\mu_p = \frac43 \mu_u - \frac13 \mu_d$, $\mu_n = \frac43 \mu_d - \frac13 \mu_u$, 
	$\mu_{\Delta^{++}} = 3\mu_u$. Using the SU(3)
	symmetry, $\mu_u=-2\mu_d = -2\mu_s$, $\mu_T$ and $F_2^T$ as well as $\mu_q$ can be written in terms of $c_1$
	or $c_2$. For example, $\mu_u = \frac23 c_1$, $\mu_T= 4 c_1$, $F_2^{\Delta^{++}} = \mu_{\Delta^{++}}-2 = 2c_1-2$.
	
	The gauge invariant nonlocal Lagrangian can be obtained 
	using the method in \cite{pingw.quantization,fw.nucleon,fw.strange}. 
	For instance, the local interaction between hyperons and $K^-$ meson can be written as
	\begin{equation}
		\mathcal{L}_{K}^{\text{local}}=
		\frac{D+F}{\sqrt{2}f}
		\bar{\Xi}^0(x) \gamma^\mu \gamma_5 \Sigma^+(x) 
		(\partial_\mu+ i e \mathscr{A}_\mu(x)) K^-(x).
	\end{equation}
	The corresponding nonlocal Lagrangian is expressed as
	\begin{equation}\label{eq:nonlocal}
	\begin{aligned}
	\mathcal L_{K}^{\text{nl}}
	&{}=
	\int \dif^4 y \frac{D+F}{\sqrt{2}f}
	\bar{\Xi}^0(x)\gamma^\mu\gamma_5 \Sigma^+(x)
	(\partial_{x,\mu} +i e \int \dif^4 a \mathscr{A}_\mu(x-a)) F(a) \\ 
	&{} \times F(x-y) \exp\left(i e \int_x^y \dif z_\nu 
	\int \dif^4 a \mathscr{A}^\nu(z-a) F(a)\right) 
	 K^-(y),
	\end{aligned}
	\end{equation}
	where $F(x)$ is the correlation function. 
	From the Lagrangian, one can see that the meson and photon fields are displaced, while the baryon fields are still at the same point. 
	In principle, we can also displace the baryon fields. As a result, the free baryon Lagrangian has to be nonlocal in order to make the total Lagrangian gauge invariant. 
	Therefore the baryon propagator and quantization will be modified. 
	The general version of this nonlocal chiral Lagrangian is much more complicated. 
	In this paper, we do not change the free Lagrangian and only the interacting Lagrangian is nonlocal. To guarantee gauge invariance, the gauge link 
	$\exp\left(i e \int_x^y \dif z_\nu \int \dif^4 a \mathscr{A}^\nu(z-a) F(a)\right) K^-(y)$ is introduced to the above Lagrangian. 
	The photon can be emitted or annihilated from the minimal substitution term or gauge link term. 
	The correlation function is associated with each photon field or meson field.
	With the correlation function, the regulator and form factors at tree level can be generated automatically at. 
	In the numerical calculation, the correlation function is chosen to be a dipole form in momentum space. 

	The nonlocal baryon-photon interaction can also be obtained in the same procedure.
	For example, the local interaction between $\Sigma^+$ and photon is written as
	\begin{equation}
		{\mathcal L}_\text{EM}^{\text{local}} =  
		-e \bar{\Sigma}^+ (x) \gamma^\mu \Sigma^+(x) \mathscr{A}_\mu(x)
		+ \frac{(c_1+3 c_2)e}{12m_\Sigma}
		 \bar{\Sigma}^+ (x) \sigma^{\mu\nu}
		\Sigma(x)^+ F_{\mu\nu}(x).
	\end{equation}
	The corresponding nonlocal Lagrangian is expressed as
	\begin{equation}
		{\mathcal L}_\text{EM}^{\text{nl}} = 
		-e \int \dif^4 a \bar{\Sigma}^+ (x) \gamma^\mu \Sigma(x)^+ \mathscr{A}_\mu(x-a)F_1(a)
		+ \frac{(c_1+3 c_2)e}{12m_\Sigma}
		\int \dif^4 a  \bar{\Sigma}^+ (x) \sigma^{\mu\nu} \Sigma(x)^+ F_{\mu\nu}(x-a)F_2(a),
	\end{equation} 
	where $F_1(a)$ and $F_2(a)$ are the correlation functions for the nonlocal electric and magnetic interactions.
	
	The momentum dependence of the form factors at tree level can be easily obtained with the 
    Fourier transformation of the correlation function. 
	As in our previous work \cite{fw.nucleon,fw.strange}, 
	the correlation function is chosen such that the charge and magnetic form factors
	at tree level have the same momentum dependence as the baryon-meson vertex, i.e. $G_M^{B,\text{tree}}(q)=\mu_B G_E^{B,\text{tree}}(q) = \mu_B \tilde{F}(q)$,
	where $\tilde{F}(q)$ is the Fourier transformation of the correlation function $F(a)$.
	Therefore, the corresponding functions $\tilde{F}_1(q)$ and $\tilde{F}_2(q)$ of $\Sigma^+$, for example, are expressed as
	\begin{equation}
		\begin{aligned}
			\tilde{F}_1^{\Sigma^+}(q)=\tilde{F}(q)
			\frac{12 {m_\Sigma}^2+(3+c_1+3 c_2) Q^2}{3(4 {m_\Sigma}^2+Q^2)}, \quad
			\tilde{F}_2^{\Sigma^+}(q)=
			\tilde{F}(q)\frac{4 (c_1+3 c_2) {m_\Sigma}^2}{3(4m_\Sigma^2+Q^2)}, 
		\end{aligned}
	\end{equation}
	where $Q^2 = -q^2$ is the momentum transfer.
	From Eq.~(\ref{eq:nonlocal}), two kinds of couplings
	between hadrons and photons can be obtained.
	One is the normal coupling, expressed as
	\begin{equation}
		{\mathcal L}^{\text{norm}} =i e \int \dif^4 y
		\frac{D+F}{\sqrt{2}f}
		\bar{\Xi}^0(x) \gamma^\mu \gamma_5 \Sigma^+(x) 
		 \int \dif^4 a \mathscr{A}_\mu(x-a) F(a) F(x-y)K^-(y).
	\end{equation}
	This interaction is similar to the traditional local Lagrangian except for the correlation function. 
	The other is the additional interaction obtained by the expansion of the gauge link, expressed as
	\begin{equation}
		{\mathcal L}^{\text{add}}=i e \int \dif^4 y 
		\frac{D+F}{\sqrt{2}f}
		\bar{\Xi}^0(x) \gamma^\mu \gamma_5 \Sigma^+(x)  
		 \partial_{x,\mu} 
		\Big(F(x-y) \int_x^y \dif z_\nu \int \dif^4 a 
		\mathscr{A}^\nu (z-a) F(a)  K^-(y) \Big). 
	\end{equation}
The additional interaction guarantees the charge conservation.
	
The Dirac and Pauli form factors of octet baryons are defined as
	\begin{equation}
		\label{eq:f1f2}
		\langle B(p^\prime)|J^\mu|B(p) \rangle=
		\bar{u}(p^\prime) \left\{ \gamma^\mu F_1^B(Q^2)+
		\frac{i\sigma^{\mu\nu} q_\nu}{2m_B}F_2^B(Q^2)
		\right\}u(p),
	\end{equation}
	where $q=p^\prime-p$.
	The electromagnetic form factors are defined as the combinations of the above form factors as
	\begin{equation}
		\begin{aligned}
			G_E^B(Q^2) {}=F_1^B(Q^2)-\frac{Q^2}{4m_B^2}F_2^B(Q^2),\quad
			G_M^B(Q^2) {}=F_1^B(Q^2) +F_2^B(Q^2).
		\end{aligned}
	\end{equation}
With the electromagnetic form factors, the magnetic and electric (charge) radii can be obtained.
The magnetic radii of octet baryons are defined as
\begin{equation}
		\langle r^2_M \rangle_B=
		\frac{-6}{G^{B}_M(0)} \frac{\dif G^{B}_M(Q^2)}{\dif Q^2} \rvert_{Q^2=0}.
\end{equation}
The electric radii of charged and neutral baryons are defined as
\begin{equation}
\langle r^2_E \rangle_B=\frac{-6}{G^{B}_E(0)} \frac{\dif G^{B}_E(Q^2)}{\dif Q^2} \rvert_{Q^2=0}
~~ \text{and} 
~~ \langle r^2_E \rangle_B=-6 \frac{\dif G^{B}_E(Q^2)}{\dif Q^2} \rvert_{Q^2=0},
\end{equation}
respectively.

\begin{figure}
	\includegraphics[width=0.8\textwidth,height=0.42\textheight]{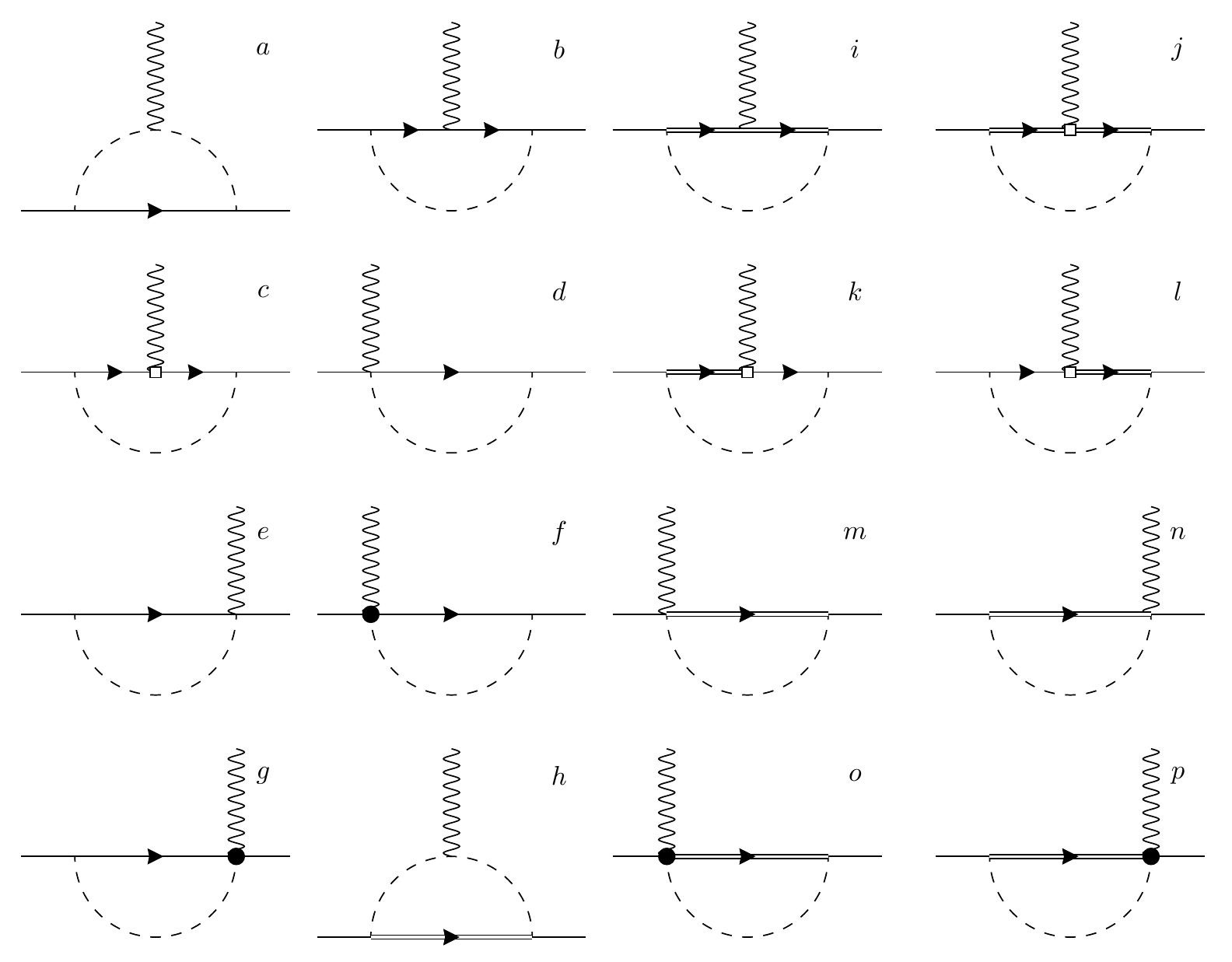}
	\caption{\label{fig.loop.diagrams} 
		One-loop Feynman diagrams for the octet electromagnetic form factors. 
		The solid, double-solid, dashed and wave lines  are for the octet baryons, 
		decuplet baryons, pseudo-scalar mesons and photons, respectively. 
		The rectangle and black dot represent magnetic and additional interacting vertex.}
\end{figure}

	According to the Lagrangian, the one loop Feynman diagrams 
	which contribute to the octet electromagnetic form factors 
	are shown in Fig.~\ref{fig.loop.diagrams}. 
From the Lagrangian, we can get the matrix element of Eq.~(\ref{eq:f1f2}). The $\pi$ meson loops have the dominant contributions, while the contributions from $K$ meson loops are much smaller due to the large $K$ meson mass. The contributions from $\eta$ and $\eta'$ loops are even smaller which are neglected in our calculation. The inclusion of these mesons does not affect the main conclusion.
The expressions for the intermediate octet and decuplet baryons are written in Appendix \ref{appendix}. In the next section, we will discuss the numerical results.

\section{Numerical Results}

The coupling constants between octet baryons and mesons are determined by the two parameters $D$ and $F$. In the numerical calculations, the parameters are chosen as $D=0.76$ and $F=0.50$ ($g_A=D+F=1.26$) \cite{Borasoy}. The coupling constant $\mathcal{C}$ is chosen to be $2D$ which is the same as in Refs.~\cite{Borasoy,Luty}. 
The off-shell parameter $z$ is $-1$ \cite{nath.decuplet}.
The physical masses are taken for mesons, octet and decuplet baryons.
The covariant regulator is chosen to be the dipole form \cite{fw.nucleon,fw.strange,Salamu.parton}
\begin{equation}
\tilde{F}(k)=\frac{\Lambda^4}{(k^2-m_j^2-\Lambda^2)^2},
\end{equation} 
where $m_j$ is the meson mass for the baryon-meson interaction and it is zero for the hadron-photon interaction. It was found that when $\Lambda$ was around 0.90 $\mathrm{GeV}$, the obtained nucleon form factors were very close to the experimental data.
Therefore, all the above parameters are predetermined. There are only two free parameters which are the low energy constants (LECs) $c_1$ and $c_2$. In our previous calculation for the nucleon form factors, they were fitted to the experimental nucleon magnetic moments \cite{fw.nucleon}. Here, $c_1$ and $c_2$ are determined to be $1.288$ and $0.420$, which give the minimal $\chi^2$ of the octet magnetic moments.

\begin{table}[]
    \small
    \begin{tabular}{c|c|c|c|c|c|c|c|c|c|c}
        \hline
        & Tree & Loop & Total & Lattice \cite{Lin} & Lattice \cite{Shanahan}& ChPT\cite{Kubis} & ChPT\cite{Blin} & NJL \cite{Serrano} & PCQM \cite{Liu} & Exp. \cite{PDG} \\ \hline
        $\mu_p$	  & 1.850 & 0.795 & $2.644\pm0.159$ & $2.4(2)$ & $2.3(3)$ & 2.61 & 2.79 & $2.78$ & $2.735\pm0.121$ & 2.793 \\ 
        $\mu_n$	  & $-0.859$ & $-1.126$ & $-1.984\pm0.216$ & $-1.59(17)$ & $-1.45(17)$ & $-1.69$ & $-1.913$ & $-1.81$ & $-1.956\pm0.103$ & $-1.913$ \\ 
        $\mu_{\Sigma^+}$  & 1.850 & 0.572 & $2.421\pm0.147$ & $2.27(16)$ & $2.12(18)$ & $2.53$ & 2.1(4) & $2.62$ & $2.537\pm0.201$ & $2.458\pm 0.010$ \\ 
        $\mu_{\Sigma^0}$	& 0.429 & 0.155 & $0.584\pm0.077$ & $-$ & $-$ & 0.76 & 0.5(2) & $-$ & $0.838\pm0.091$ & $-$ \\ 
        $\mu_{\Sigma^-}$  & $-0.991$ & $-0.262$ & $-1.253\pm0.008$ & $-0.88(8)$ & $-0.85(10)$ & $-1.00$ & $-1.1(1) $ & $-1.62$ & $-0.861\pm0.040$ & $-1.160\pm0.025$ \\ 
        $\mu_\Lambda$	  & $-0.429$ & $-0.165$ & $-0.594\pm0.057$ & $-$ & $-$  & $-0.76$ & $-0.5(2)$ & $-$ & $-0.867\pm0.074$ & $-0.613\pm0.004$ \\ 
        $\mu_{\Xi^0}$  & $-0.859$ & $-0.521$ & $-1.380\pm0.169$ & $-1.32(4)$ & $-1.07(7) $ & $-1.51$ & $-1.0(4)$ & $-1.14$ & $-1.690\pm0.142$ & $-1.250\pm0.014$ \\ 
        $\mu_{\Xi^-}$  & $-0.991$ & 0.266 & $-0.725\pm0.077$ & $-0.71(3)$ & $-0.57(5)$ & $-0.93$ & $-0.7(1)$ & $-0.67$ & $-0.840\pm0.087$ & $-0.651\pm0.080$ \\ 
        \hline
    \end{tabular}
    \caption{The tree, loop and total contributions to the octet magnetic moments $\mu_B$ (in units of the nucleon magneton $\mu_N$). The results from two lattice simulations, ChPT with IR and EOMS scheme, NJL and PCQM models as well as the experimental data are also listed.}
    \label{Mag}
\end{table}

In Table \ref{Mag}, the tree, loop and total contributions to the baryon magnetic moments obtained from the nonlocal chiral effective theory are listed. The wave-function renormalization constant $Z$ is included in the calculation, i.e., tree-level contribution has been multiplied by $Z$. 
The error bar in our calculation is determined by varying $\Lambda$ from 0.8 to 1.0 GeV. The results from two lattice simulations \cite{Boinepalli,Shanahan}, ChPT with IR \cite{Kubis} and EOMS scheme \cite{Blin}, NJL and PCQM models \cite{Serrano,Liu} as well as the experimental data are also listed for comparison. From the table, one can see that all the magnetic moments of octet baryons are reasonably reproduced. The largest deviation from the experiments is for $\Xi$s, where the calculated central values of magnetic moments of $\mu_{\Xi^0}$ and $\mu_{\Xi^-}$ are about $10\%$ larger than experimental ones. For the other baryons, the deviation from the experiments is less than $5\%$. Considering the error bar, the calculated magnetic moments of octet baryons are in very good agreement with the experimental values. It is interesting that all the tree and loop contributions have the same signs except for $\Xi^-$, where the loop diagrams give the opposite contribution to the tree diagram. The data from lattice simulations are somewhat smaller which is partially due to the large pion mass and/or the neglecting of the disconnected contribution. 
The results at order of ${\cal O}(p^3)$ in ChPT with IR are listed for comparison, where the calculated moments of most baryons are comparable with the experimental data. The magnetic moment of $\Xi^-$ is about $40-50\%$ larger which could be decreased by the inclusion of intermediate decuplet states. At order of ${\cal O}(p^4)$, with the additional five low energy constants, the seven exerimental magnetic moments can be exactly reproduced \cite{Kubis}. 
For the ChPT with EOMS scheme, the results with the inclusion of intermediate decuplet states and vector mesons are listed. It was found that the inclusion of intermediate decuplet 
states improves the results, especially for $\Xi^-$ and makes the results at ${\cal O}(p^3)$ as good as those at ${\cal O}(p^4)$ \cite{Blin}. Our calculation in nonlocal EFT comfirms that the results at one-loop level with the inclusion of decuplet states are good enough to reproduce the experimental values.

\begin{figure}[htbp]
\begin{center}
\includegraphics[width=.6\textwidth]{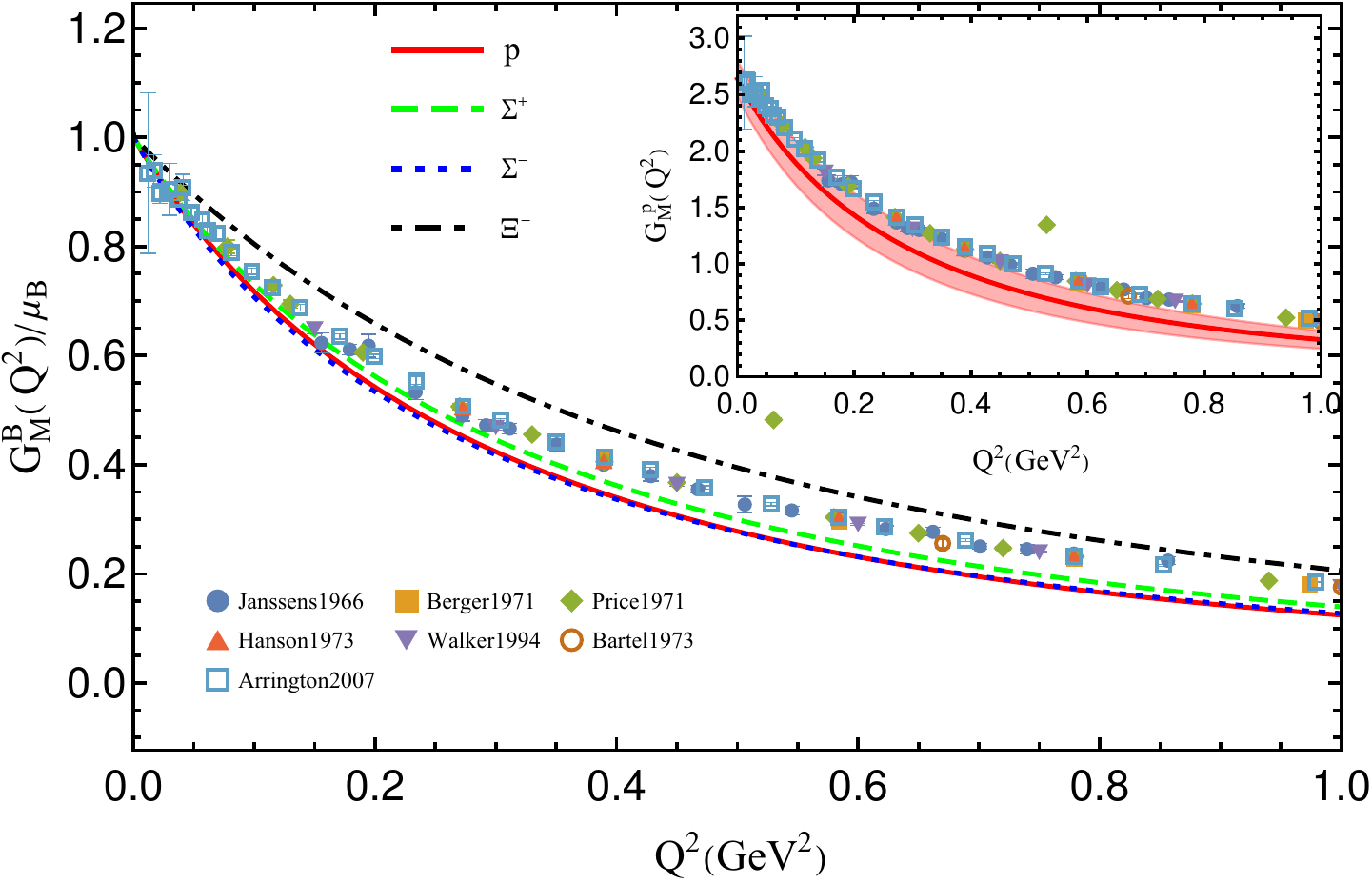}
\caption{The normalized magnetic form factors of charged octet baryons $G_M^B/\mu_B$ versus momentum transfer $Q^2$. The solid, dashed, dotted and dash-dotted lines are for proton, $\Sigma^+$, $\Sigma^-$ and $\Xi^-$, respectively. The magnetic form factor of proton with $\Lambda$ varying from 0.8 to 1 GeV is also plotted at the corner of the figure. The experimental form factor of proton is from Refs.~\cite{
	Janssens1966,Berger1971,Price1971,Anklin1994,Walker1994,Bartel1973,Arrington2007}.} 
\label{magc}
\end{center}
\end{figure}

\begin{figure}[htbp]
\begin{center}
\includegraphics[width=.6\textwidth]{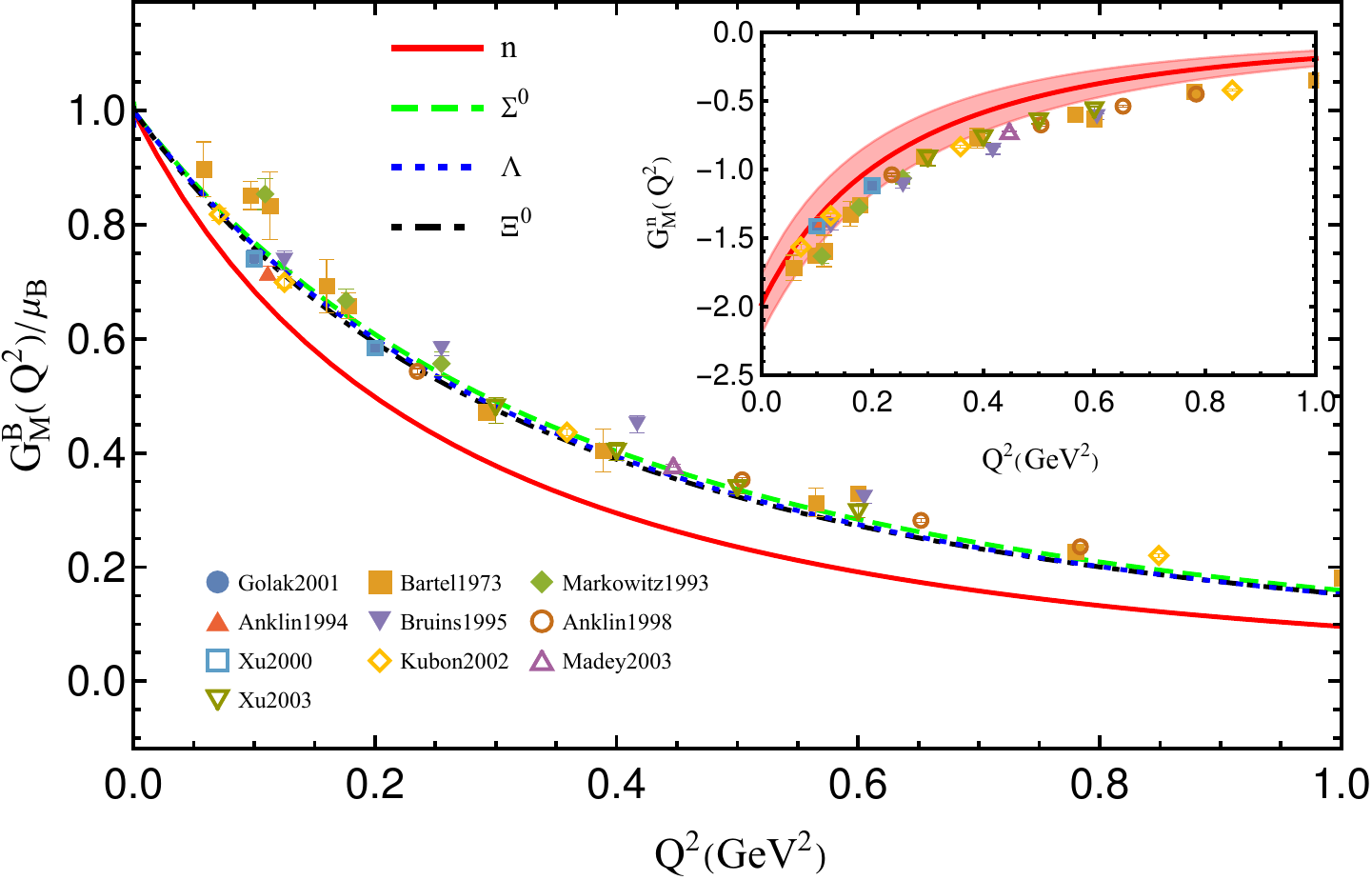}
\caption{The normalized magnetic form factors of neutral octet baryons $G_M^B/\mu_B$ versus momentum transfer $Q^2$. The solid, dashed, dotted and dash-dotted lines are for neutron, $\Sigma^0$, $\Lambda$ and $\Xi^0$, respectively. The magnetic form factor of the neutron with $\Lambda$ varying from 0.8 to 1 GeV is also plotted at the corner of the figure. The experimental form factor of the neutron is from Refs.~\cite{
Golak2001,Bartel1973,Markowitz1993,Anklin1994,Bruins1995,
Anklin1998,Xu2000,Kubon2002,Madey2003,Xu2003}.}
\label{magn}
\end{center}
\end{figure}

The magnetic form factors of charged octet baryons versus momentum transfer $Q^2$ are plotted in Fig. \ref{magc}. The solid, dashed, dotted and dash-dotted lines are for proton, $\Sigma^+$, $\Sigma^-$ and $\Xi^-$, respectively. The magnetic form factor of proton with $\Lambda$ varying from 0.8 to 1 GeV is also plotted at the corner of the figure. Considering the error bar, it is clear the proton magnetic form factor is comparable with the experimental data up to 1 GeV$^2$. This is the advantage of the nonlocal chiral effective theory. The correlation function in the nonlocal Lagrangian makes the loop integral convergent. In the mean time, it provides the momentum dependence of the form factors at tree level, and as a result, the total form factors can be close to the experimental data up to relatively large $Q^2$. The normalized magnetic form factors of the nucleon were studied in the chiral perturbation theory including $\rho$ and $\omega$ mesons as well as the $\Delta$ resonance \cite{Bauer}.  Two of the undetermined low energy coupling constants were adjusted to the nucleon magnetic moments while the remaining six LECs were fitted simultaneously to the experimental data up to $Q^2 = 0.4$ GeV$^2$. It was found that the results incorporating vector mesons agree well with experimental data in a momentum transfer region $0 \leq Q^2 \leq 0.4$ GeV$^2$.
The other form factors of charged baryons have a similar momentum dependence as proton. Among them, $\Xi^-$ decreases a little slower with increasing $Q^2$. The magnetic radii are determined by the slopes of the form factors at zero momentum transfer which will be discussed later. 

The normalized magnetic form factors for the charge neutral baryons are plotted in Fig. \ref{magn}. The solid, dashed, dotted and dash-dotted lines are for neutron, $\Sigma^0$, $\Lambda$ and $\Xi^0$, respectively. The band in the small figure is for the magnetic form factor of the neutron with $\Lambda$ varying from 0.8 to 1 GeV. The magnetic from factors of $\Sigma^0$, $\Lambda$ and $\Xi^0$ are close to each other. The normalized neutron magnetic form factor is a little smaller than the experimental data and it drops faster than the other three neutral baryons. Taking the error bar into account, the calculated neutron magnetic form factor is still close to the experiments. The smaller normalized magnetic form factor of the neutron is partially because of its larger calculated moment. All of the form factors of octet baryons have a dipole-like momentum dependence.

The tree, loop and total contributions to the magnetic radii of octet baryons are listed in Table \ref{radiiM}. The data from lattice simulation, chiral perturbation theory and phenomenological quark models are also listed for comparison. Though our central values for proton and neutron are a little larger than experiments, the results are still reasonable. The magnetic radii of octet baryons vary from 0.5 fm$^2$ to 0.9 fm$^2$, but show no simple dependence on baryon/quark mass. $\Xi^-$ has the largest contribution at tree level. Because of the opposite contribution from the loop diagrams, its total radius is the smallest one. Amazing thing is though the values from different methods are quite different, the order of the values from the largest to smallest is almost the same. For example, $\Sigma^-$ and $\Xi^-$ have the largest and smallest magnetic radii, respectively. The neutron magnetic radius is the second largest one. Our results also show the tree and loop contributions are strongly baryon dependent. The loop contribution to $\langle r_M^2 \rangle_{\Sigma^0}$ is less than one half of the tree contribution. However, for neutron, the loop contribution is twice bigger than the tree contribution. 

\begin{table}[htbp]
	\small
	\begin{tabular}{c|c|c|c|c|c|c|c|c|c|c}
\hline
 & Tree & Loop & Total & Lattice \cite{Boinepalli} & Lattice \cite{Shanahan} & ChPT\cite{Kubis} & ChPT\cite{Blin}& NJL \cite{Serrano} & PCQM \cite{Liu} & Exp. \cite{PDG} \\ \hline
$\langle r_M^2 \rangle_p$	  & 0.403 & 0.382 & $0.785\pm0.132$ & $0.470(48)$ & $0.71(8)$ & 0.699 & 0.9(2) & $0.76$ & $0.909\pm0.084$ & $0.72\pm0.04$ \\ 
$\langle r_M^2 \rangle_n$	  & 0.250 & 0.596 & $0.845\pm0.148$ & $0.478(50)$ & $0.86(9)$ & 0.790 & 0.8(2) & $0.83$ & $0.922\pm0.079$ & $0.75\pm 0.02$ \\ 
$\langle r_M^2 \rangle_{\Sigma^+}$  & 0.441 & 0.324 & $0.765\pm0.131$ & $0.466(42)$ & $0.66(5)$ & $0.80\pm 0.05$ & 1.2(2) & $0.77$ & $0.885\pm0.094$ & $-$ \\ 
$\langle r_M^2 \rangle_{\Sigma^0}$	& 0.424 & 0.194 & $0.618\pm0.124$ & $0.432(38)$ & $-$ & $0.45\pm 0.08 $ & 1.1(2) & $-$ & $0.851\pm0.102$ & $-$ \\ 
$\langle r_M^2 \rangle_{\Sigma^-}$	  & 0.456 & 0.445 & $0.901\pm0.119$ & $0.483(49)$ & $1.05(9)$ & $1.20\pm 0.13 $ & 1.2(2) & $0.92$ & $0.951\pm0.083$ & $-$ \\ 
$\langle r_M^2 \rangle_\Lambda$	  & 0.417 & 0.203 & $0.620\pm0.126$ & $0.347(24)$ & $-$ & $0.48\pm 0.09$ & 0.6(2) & $-$ & $0.852\pm0.103$ & $-$ \\ 
$\langle r_M^2 \rangle_{\Xi^0}$  & 0.359 & 0.298 & $0.657\pm0.128$ & $0.384(22)$ & $0.53(5)$ & $0.61\pm 0.12$ & 0.7(3) & $0.44$ & $0.871\pm0.099$ & $-$ \\ 
$\langle r_M^2 \rangle_{\Xi^-}$  & 0.789 & $-0.255$ & $0.534\pm0.135$ & $0.336(18)$ & $0.44(5)$ & $0.50\pm 0.16$ & 0.8(1) & $0.26$ & $0.840\pm0.109$ & $-$ \\ 
\hline
	\end{tabular}
\caption{The tree, loop and total contributions to the octet magnetic radii $\langle r_M^2 \rangle_B$ (in units of fm$^2$). The results from two lattice simulations, ChPT with IR and EOMS scheme, NJL and PCQM models as well as the experimental data are also listed.}
	\label{radiiM}
\end{table}

\begin{figure}[htbp]
	\begin{center}
		\includegraphics[width=.6\textwidth]{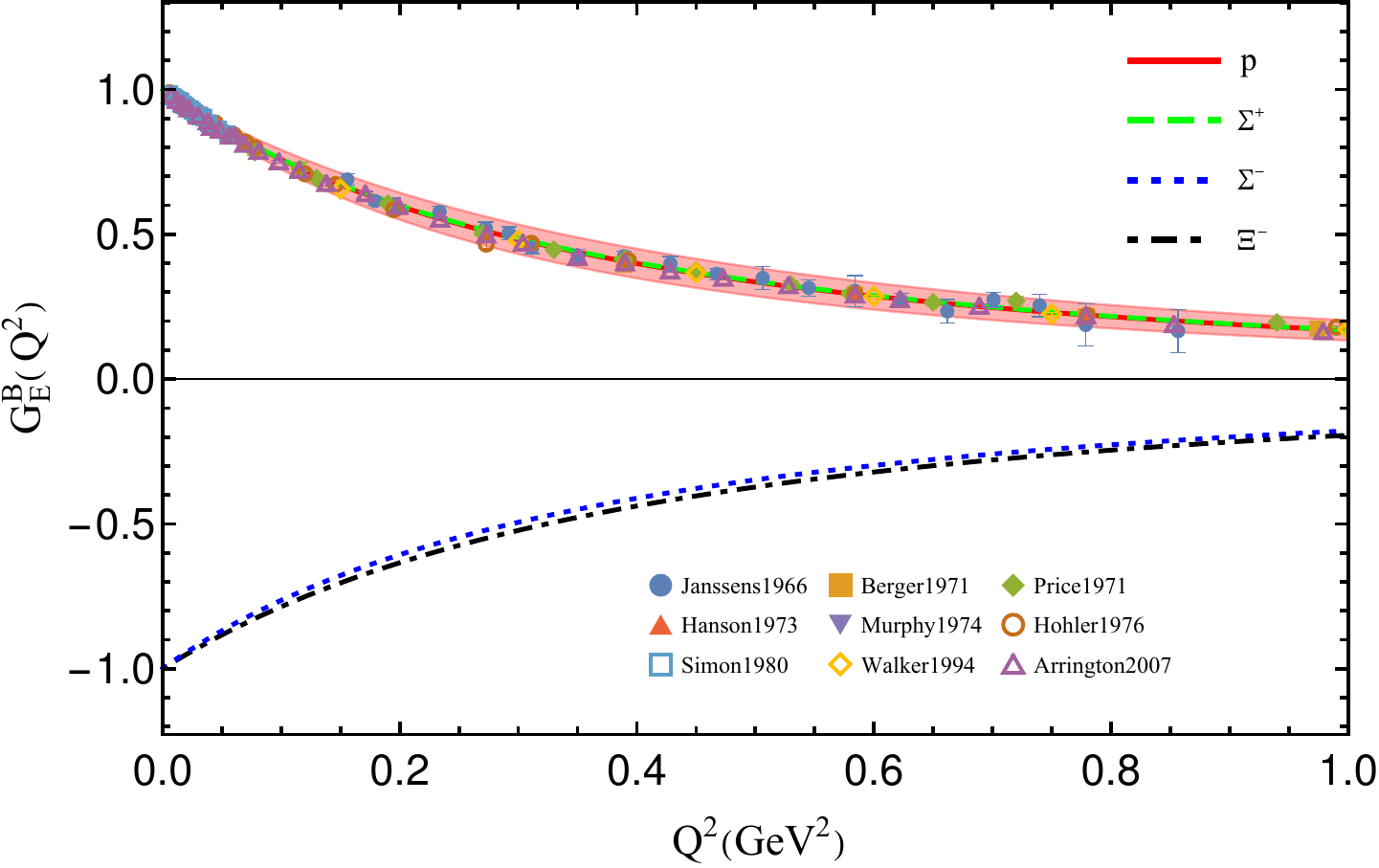}
		\caption{Same as Fig. 2 but for electric form factors, 
		The experimental form factor of proton is from Refs.~\cite{
			Janssens1966,Berger1971,Price1971,Hanson1973,Murphy1974,
			Hohler1976,Simon1980,Walker1994,Arrington2007}.}
		\label{chargec}
	\end{center}
\end{figure}

We now discuss the electric form factors. Similar as for magnetic form factors, the bands are also shown for nucleon electric form factors with $0.8$ GeV $\leq \Lambda \leq 1$ GeV. In Fig.~\ref{chargec}, we plot the electric form factors of the charged baryons. Because of the additional interaction which makes the nonlocal Lagrangian locally gauge invariant, the electric form factors start from their charge at $Q^2=0$. The proton charge form factor is close to the experimental data. The absolute values of the electric form factors of charged baryons have a similar momentum dependence. This could be examined by the further experiments and/or accurate lattice simulation.

\begin{figure}[htbp]
	\begin{center}
		\includegraphics[width=.6\textwidth]{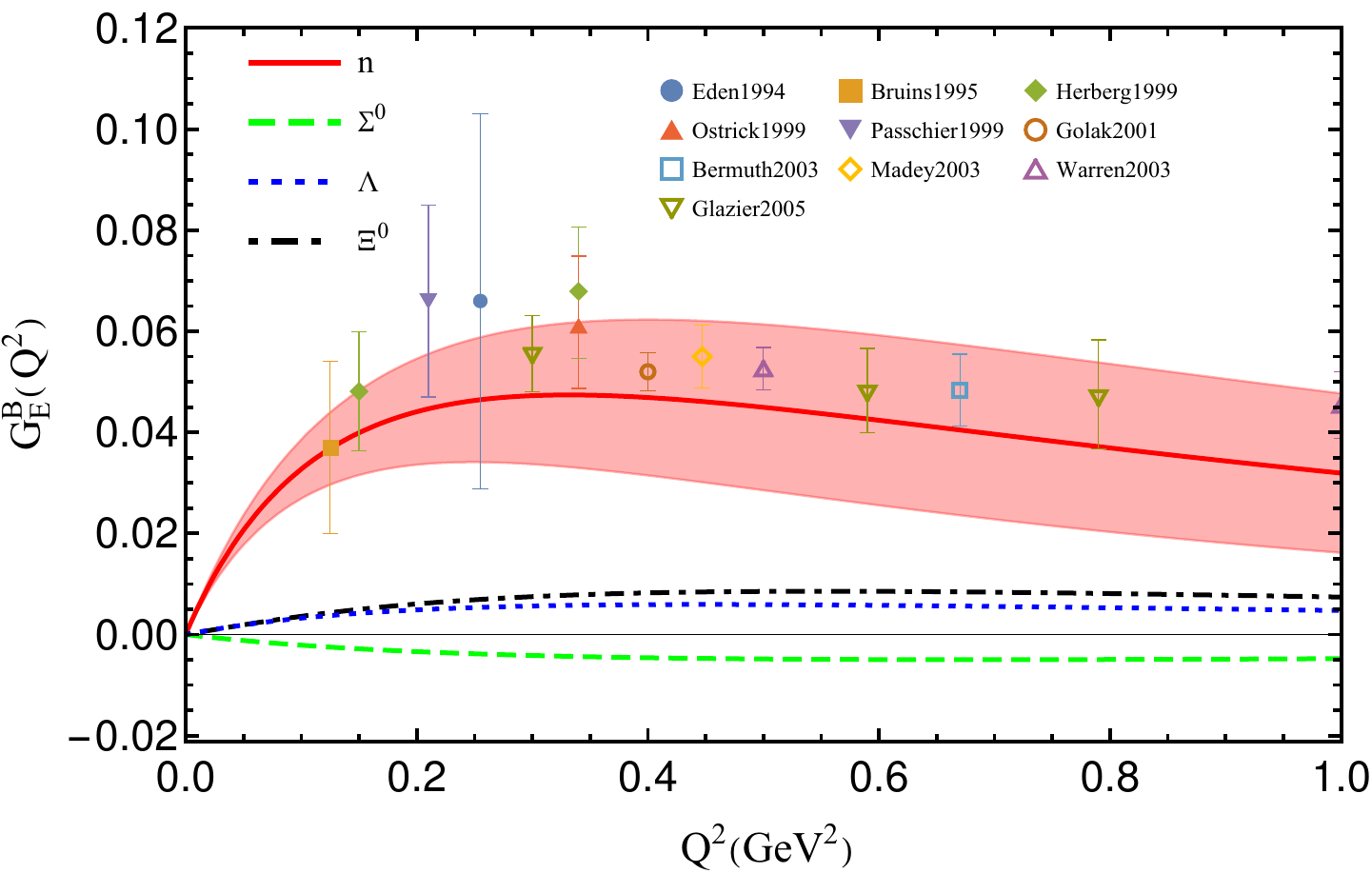}
		\caption{Same as Fig. 3 but for electric form factors, 
			The experimental form factor of the neutron is from Refs.~\cite{
				Eden1994,Bruins1995,Herberg1999,Ostrick1999,Passchier1999,
				Golak2001,Bermuth2003,Madey2003,Warren2003,Glazier2005}.}
		\label{chargen}
	\end{center}
\end{figure}

The electric form factors for the neutral baryons are plotted in Fig.~\ref{chargen}. Again due to the charge conservation, the form factors start from 0 at zero momentum transfer. The calculated electric form factor of the neutron is consistent with the experimental data. The form factors of the other neutral baryons are very small. There is no tree level contribution to the electric form factors of neutral baryons and all the contributions are from the loop diagrams. Among them, the neutron has the largest contribution from $\pi$-loop diagrams. The corresponding $\pi$-loop diagrams for the other neutral baryons are fairy small due to the small coupling constants.

The charge radii of octet baryons are listed in Table~\ref{radiiC}. Our results are comparable with the experimental data in PDG for nucleon and $\Sigma^-$. A small proton charge radius $\langle r_E \rangle_p=0.831\pm0.007\pm0.012$, i.e. $\langle r_E^2 \rangle_p=0.691\pm0.032$ was reported recently \cite{Xiong} which is also close to our value $\langle r_E^2 \rangle _p=0.729\pm0.112$. For the neutral baryons, the loop contribution is very small except that of the neutron. For the charged baryons, the tree level contributions are the same which are also dominant for all of them. The loop contribution has the same order of magnitude except for $\Xi^-$, where the loop contribution is small. Different from the magnetic radii, the total charge radii vary around 0.6 and 0.7 fm$^2$ for the charged baryons.
Though the charge radii of charged baryons from different models are comparable, the predictions for neutral baryons (both sign and size) are quite different.

\begin{table}[htbp]
	\footnotesize
	\begin{tabular}{c|c|c|c|c|c|c|c|c|c|c}
		\hline
 & Tree & Loop & Total & Lattice\cite{Wang} & Lattice\cite{Shanahan2} & ChPT\cite{Kubis} & ChPT\cite{Blin} & NJL\cite{Serrano} & PCQM \cite{Liu} & Exp. \cite{PDG} \\ \hline
$\langle r_E^2\rangle_p$	  & 0.577 & 0.152 & $0.729\pm0.112$ & $0.685(66)$ & $0.76(10)$ & 0.717 & 0.878 & $0.76$ & $0.767\pm0.113$ & $0.707\pm0.0007$ \\ 
$\langle r_E^2 \rangle_n$	  & 0 & $-0.146$ & $-0.146\pm0.018$ & $-0.158(33)$ & $-$ & $-0.113$ & $0.03(7)$ & $-0.14$ & $-0.014\pm0.001$ & $-0.116\pm0.0022$ \\ 
$\langle r_E^2 \rangle_{\Sigma^+}$  & 0.577 & 0.142 & $0.719\pm0.116$ & $0.749(72)$ & $0.61(8)$ & $0.60\pm 0.02$ & 0.99(3) & $0.92$ & $0.781\pm0.108$ & $-$ \\ 
$\langle r_E^2 \rangle_{\Sigma^0}$	& 0 & 0.010 & $0.010\pm0.004$ & $-$ & $-$ & $-0.03\pm 0.01$ & 0.10(2) & $-$ & $0$ & $-$ \\ 
$\langle r_E^2 \rangle_{\Sigma^-}$	  & 0.577 & 0.123 & $0.700\pm0.124$ & $0.657(58)$ & $0.45(3)$ & $0.67\pm 0.03$ & 0.780 & $0.74$ & $0.781\pm0.063$ & $0.61\pm0.16$ \\ 
$\langle r_E^2 \rangle_\Lambda$	  & 0 & $-0.015$ & $-0.015\pm0.004$ & $0.010(9)$ & $-$ & $0.11\pm 0.02$ & 0.18(1) & $-$ & $0$ & $-$ \\ 
$\langle r_E^2 \rangle_{\Xi^0}$  & 0 & $-0.015$ & $-0.015\pm0.007$ & $0.082(29)$ & $-$ & $0.13\pm 0.03$ & 0.36(2) & $0.24$ & $0.014\pm0.008$ & $-$ \\ 
$\langle r_E^2 \rangle_{\Xi^-}$  & 0.577 & 0.025 & $0.601\pm0.127$ & $0.502(47)$ & $0.37(2)$ & $0.49\pm 0.05$ & 0.61(1) & $0.58$ & $0.767\pm0.113$ & $-$ \\ 
\hline
	\end{tabular}
\caption{The tree, loop and total contributions to the octet charge radii $\langle r_E^2 \rangle_B$ (in units of fm$^2$). The results from two lattice simulations, ChPT with IR and EOMS scheme, NJL and PCQM models as well as the experimental data are also listed.}
	\label{radiiC}
\end{table}

\section{Summary}
We applied the nonlocal chiral effective theory to study the electromagnetic form factors of octet baryons. The correlation function in the Lagrangian makes the loop integral convergent. It also provides the momentum dependence of the form factors at tree level. The additional interaction generated from the expansion of the gauge link guarantees the Lagrangian is locally gauge invariant. This nonlocal Lagrangian makes it possible to study the physical quantities at relatively large momentum transfer in the framework of chiral effective theory. In the numerical calculation, all the parameters are predetermined except the two low energy constants $c_1$ an $c_2$. They are fitted to give the minimum of $\chi^2$ of the octet magnetic moments. When extending the previous study of form factors of nucleons to all the octet baryons, we do not add any new parameter. The magnetic moments are well reproduced. The deviation from the experiments is less than $5\%$ except $\Xi^0$ and $\Xi^-$, where the deviation of the central value is about $10\%$. For the radii, most experiments focus on the nucleon and there is few data for the other baryons. Considering the error bar, all our results on magnetic moments and electromagnetic radii are in very good agreement with the current experimental data. The calculated nucleon form factors are close to the experiments up to $Q^2=1$ GeV$^2$. For the other octet baryons, since the method is the same, we expect this nonlocal Lagrangian can also give good descriptions. The difference between our results and those of other theoretical methods could be examined by future experiments and more accurate lattice simulations.

\section*{Acknowledgments}
This work is supported by the National Natural Sciences Foundations of China under the grant No. 11975241,
the Sino-German CRC 110 ``Symmetries and the Emergence of Structure in QCD" project by NSFC under the grant
No.11621131001.

\appendix

\section{Loop Expressions}

\label{appendix}
	
In this section, we show the expressions of loop integrals for the intermediate octet and decuplet baryons. 
Let's take the $\Sigma$ hyperons as an example. 

The contributions of Fig.~\ref{fig.loop.diagrams}a are written as
	\begin{align}
		\Gamma_a^\mu(\Sigma^-)&{}=
		\frac{F^2}{f^2}
		I_a^{\mu,\Sigma \pi}
		%%%%%%%%%%%%
		+
		\frac{D^2}{3 f^2}
		I_a^{\mu,\Lambda \pi}
		%%%%%%%%%%%%
		+
		\frac{(D-F)^2}{2 f^2}
		I_a^{\mu,N K} , \\
		%%%%%%%%%++++++++++++++++++
		\Gamma_a^\mu(\Sigma^0)&{}=
		\frac{(D-F)^2}{4 f^2}
		I_a^{\mu,N K}
		%%%%%%%%%%%%
		-\frac{(D+F)^2}{4 f^2}
		I_a^{\mu,\Xi K} ,\\
		%%%%%%%%%%%%
		%%%%%%%%%++++++++++++++++++
		\Gamma_a^\mu(\Sigma^+)&{}=
		-\frac{F^2}{f^2}
		I_a^{\mu,\Sigma \pi}
		%%%%%%%%%%%%
		-\frac{D^2}{3 f^2}
		I_a^{\mu,\Lambda \pi}
		%%%%%%%%%%%%
		-\frac{(D+F)^2}{2 f^2}
		I_a^{\mu,\Xi K},
	\end{align} 
	where the integral $I_a^{\mu,BM}$ is expressed as
	\begin{equation} 
		I_a^{\mu,BM}=
		\bar{u}(p^\prime)\tilde F(q)\int \frac{\dif^4 k}{(2\pi)^4}
		\frac{\tilde F(q+k)\tilde F(k)}{D_M(k+q)}
		\frac{-(2k+q)^\mu}{D_M(k)}
		(\slashed{k}+\slashed{q})\gamma_5
		\frac{1}{\slashed{p}-\slashed{k}-m_B}
		\slashed{k}\gamma_5 u(p).
	\end{equation}
	$D_M(k)$ is defined as
	\begin{equation} 
		D_M(k)=k^2-m_M^2+i\epsilon.
	\end{equation}
	$m_B$ and $m_M$ are the masses of the intermediate $B$ baryon and $M$ meson. 

The contributions of Fig.~\ref{fig.loop.diagrams}b are expressed as
		\begin{align}
			\Gamma_b^\mu(\Sigma^-) {}&{}=
			\frac{F \left(c_1 D Q^2-F \left(\left(-2 c_1+3 c_2+3\right) Q^2+12 m_{\Sigma }^2\right)\right)}{3 f^2 \left(4 m_{\Sigma }^2+Q^2\right)}
			I_b^{\mu,\Sigma \pi}
			%%%%%%%%%%%%
			-\frac{c_1 D Q^2 (D-3 F)}{9 f^2 \left(4 m_{\Lambda }^2+Q^2\right)}
			I_b^{\mu,\Lambda \pi} 
			%%%%%%%%%%%%
			\\ \nonumber {}&{}
			-\frac{c_1 Q^2 (D-F)^2}{3 f^2 \left(4 m_N^2+Q^2\right)}
			I_b^{\mu,N K}
			%%%%%%%%%%%%
			-\frac{(D+F)^2 \left(\left(-c_1+3 c_2+3\right) Q^2+12 m_{\Xi }^2\right)}{6 f^2 \left(4 m_{\Xi }^2+Q^2\right)}
			I_b^{\mu,\Xi K}, \\
			%%%%%%%%%%%%+++++++++++
			\Gamma_b^\mu(\Sigma^0) 
			&{}=
			\frac{2 c_1 F^2 Q^2}{3 f^2 \left(4 m_{\Sigma }^2+Q^2\right)}
			I_b^{\mu,\Sigma \pi}
			%%%%%%%%%%%%
			-\frac{c_1 D^2 Q^2}{9 f^2 \left(4 m_{\Lambda }^2+Q^2\right)}
			I_b^{\mu,\Lambda \pi} 
			\\ \nonumber {}&{}
			%%%%%%%%%%%%
			+\frac{(D-F)^2 \left(\left(-c_1+3 c_2+3\right) Q^2+12 m_N^2\right)}{12 f^2 \left(4 m_N^2+Q^2\right)}
			I_b^{\mu,N K}
			%%%%%%%%%%%%
			-\frac{(D+F)^2 \left(\left(c_1+3 c_2+3\right) Q^2+12 m_{\Xi }^2\right)}{12 f^2 \left(4 m_{\Xi }^2+Q^2\right)}
			I_b^{\mu,\Xi K}, \\
	%%%%%%%%%%%%+++++++++++
			\Gamma_b^\mu(\Sigma^+) {}&{}=
			\frac{F \left(c_1 (-D) Q^2+\left(2 c_1+3 c_2+3\right) F Q^2+12 F m_{\Sigma }^2\right)}{3 f^2 \left(4 m_{\Sigma }^2+Q^2\right)}
			I_b^{\mu,\Sigma \pi}
			%%%%%%%%%%%%
			-\frac{c_1 D Q^2 (D+3 F)}{9 f^2 \left(4 m_{\Lambda }^2+Q^2\right)}
			I_b^{\mu,\Lambda \pi}
			%%%%%%%%%%%%
			\\  \nonumber {}&{}
			+\frac{(D-F)^2 \left(\left(c_1+3 c_2+3\right) Q^2+12 m_N^2\right)}{6 f^2 \left(4 m_N^2+Q^2\right)}
			I_b^{\mu,N K}
			%%%%%%%%%%%%
			-\frac{c_1 Q^2 (D+F)^2}{3 f^2 \left(4 m_{\Xi }^2+Q^2\right)}
			I_b^{\mu,\Xi K},
		\end{align}
	where the integral $I_b^{\mu,BM}$ is written as
	\begin{equation} 
		I_b^{\mu,BM}=
		\bar{u}(p^\prime)\tilde F(q)\int \frac{\dif^4 k}{(2\pi)^4}
		\frac{\tilde F(k)^2}{D_M(k)}
		\slashed{k}\gamma_5 
		\\ \frac{1}{\slashed{p'}-\slashed{k}-m_B}\gamma^\mu 
		\frac{1}{\slashed{p}-\slashed{k}-m_B}
		\slashed{k}\gamma_5
		u(p).
	\end{equation}
	
Fig.~\ref{fig.loop.diagrams}c is similar to Fig.~\ref{fig.loop.diagrams}b except for the magnetic interaction. 
	The contributions of this diagram are written as
		\begin{align}
			\Gamma_c^\mu(\Sigma^-) {}&{}=
			\frac{2  F m_{\Sigma } \left(c_1 (D+2 F)-3 c_2 F\right)}{3 f^2 \left(4 m_{\Sigma }^2+Q^2\right)}
			I_c^{\mu,\Sigma \pi}
			%%%%%%%%%%%%
			-\frac{2  c_1 D (D-3 F) m_{\Lambda }}{9 f^2 \left(4 m_{\Lambda }^2+Q^2\right)}
			I_c^{\mu,\Lambda \pi} 
			%%%%%%%%%%%%
			\\  \nonumber {}&{}
			-\frac{2  c_1 (D-F)^2 m_N}{3 f^2 \left(4 m_N^2+Q^2\right)}
			I_c^{\mu,N K}
			%%%%%%%%%%%%
			%%\\  &{}
			+ \frac{ \left(c_1-3 c_2\right) (D+F)^2 m_{\Xi }}{3 f^2 \left(4 m_{\Xi }^2+Q^2\right)}
			I_c^{\mu,\Xi K}, \\
	%%%%%%%%%%%%+++++++++++
			\Gamma_c^\mu(\Sigma^0) 	{}&{} =
			\frac{4  c_1 F^2 m_{\Sigma }}{3 f^2 \left(4 m_{\Sigma }^2+Q^2\right)}
			I_c^{\mu,\Sigma \pi}
			%%%%%%%%%%%%
			-\frac{2  c_1 D^2 m_{\Lambda }}{9 f^2 \left(4 m_{\Lambda }^2+Q^2\right)}
			I_c^{\mu,\Lambda \pi} 
			\\ \nonumber {}&{}
			%%%%%%%%%%%%
			-\frac{ \left(c_1-3 c_2\right) (D-F)^2 m_N}{6 f^2 \left(4 m_N^2+Q^2\right)}
			I_c^{\mu,N K}
			%%%%%%%%%%%%
			-\frac{ \left(c_1+3 c_2\right) (D+F)^2 m_{\Xi }}{6 f^2 \left(4 m_{\Xi }^2+Q^2\right)}
			I_c^{\mu,\Xi K}, \\
	%%%%%%%%%%%%+++++++++++
			\Gamma_c^\mu(\Sigma^+) {}&{}=
			\frac{2  F m_{\Sigma } \left(c_1 (-D)+2 c_1 F+3 c_2 F\right)}{3 f^2 \left(4 m_{\Sigma }^2+Q^2\right)}
			I_c^{\mu,\Sigma \pi}
			%%%%%%%%%%%%
			-\frac{2  c_1 D (D+3 F) m_{\Lambda }}{9 f^2 \left(4 m_{\Lambda }^2+Q^2\right)}
			I_c^{\mu,\Lambda \pi}
			%%%%%%%%%%%%
			\\  \nonumber {}&{}
			+ \frac{ \left(c_1+3 c_2\right) (D-F)^2 m_N}{3 f^2 \left(4 m_N^2+Q^2\right)}
			I_c^{\mu,N K}
			%%%%%%%%%%%%
			-\frac{2  c_1 (D+F)^2 m_{\Xi }}{3 f^2 \left(4 m_{\Xi }^2+Q^2\right)}
			I_c^{\mu,\Xi K},
		\end{align}
	where $I_c^{\mu,B M}$ is expressed as
	\begin{equation}
		I_c^{\mu,BM}=
		\bar{u}(p^\prime)\tilde F(q)\int \frac{\dif^4 k}{(2\pi)^4}
		\frac{\tilde F(k)^2}{D_M(k)}
		\slashed{k}\gamma_5
		\frac{1}{\slashed{p'}-\slashed{k}-m_B} \\
		i \sigma^{\mu\nu}q_{\nu}
		\frac{1}{\slashed{p}-\slashed{k}-m_B}
		\slashed{k}\gamma_5
		u(p).
	\end{equation}
	
	Figs.~\ref{fig.loop.diagrams}d and \ref{fig.loop.diagrams}e are the Kroll-Ruderman diagrams. The contributions of these two diagrams are written as
	\begin{align}
		\Gamma_{d+e}^\mu(\Sigma^-) {}&=
		\frac{F^2}{f^2}
		I_{d+e}^{\mu,\Sigma \pi}
		%%%%%%%%%%%%
		+\frac{D^2}{3 f^2}
		I_{d+e}^{\mu,\Lambda \pi} 
		%%%%%%%%%%%%
		+\frac{(D-F)^2}{2 f^2}
		I_{d+e}^{\mu,N K} , \\
		%%%%++++++++++++++++
		%%%%++++++++++++++++
		\Gamma_{d+e}^\mu(\Sigma^0) {}&=
		\frac{(D-F)^2}{4 f^2}
		I_{d+e}^{\mu,N K}
		%%%%%%%%%%%%
		-\frac{(D+F)^2}{4 f^2}
		I_{d+e}^{\mu,\Xi K},  \\
		%%%%++++++++++++++++
		%%%%++++++++++++++++
		\Gamma_{d+e}^\mu(\Sigma^+) {}&=
		-\frac{F^2}{f^2}
		I_{d+e}^{\mu,\Sigma \pi}
		%%%%%%%%%%%%
		-\frac{D^2}{3 f^2}
		I_{d+e}^{\mu,\Lambda \pi} 
		%%%%%%%%%%%%
		-\frac{(D+F)^2}{2 f^2}
		I_{d+e}^{\mu,\Xi K}, 
	\end{align} 
	where
	\begin{equation}
		\begin{aligned}
			I_{d+e}^{\mu,BM}{}&{}=
			\bar{u}(p^\prime)\tilde F(q)\int \frac{\dif^4 k}{(2\pi)^4}
			\frac{\tilde{F}(k)^2}{D_M(k)}
			\Big\lbrace
			\slashed{k}\gamma_5
			\frac{1}{\slashed{p'}-\slashed{k}-m_B} 
			%% \\ & 
			\gamma^\mu\gamma_5 
			+
			\gamma^\mu\gamma_5
			\frac{1}{\slashed{p}-\slashed{k}-m_B} 
			\slashed{k}\gamma_5
			 \Big\rbrace u(p).
		\end{aligned}
	\end{equation} 
	
	Figs.~\ref{fig.loop.diagrams}f and \ref{fig.loop.diagrams}g are the additional diagrams which are generated from the expansion of the gauge link terms.
	The contributions of these two diagrams for intermediate octet hyperons are expressed as
	\begin{align}
		\Gamma_{f+g}^\mu(\Sigma^-){}&{}=
		\frac{F^2}{f^2}
		I_{f+g}^{\mu,\Sigma \pi}
		%%%%%%%%%%%%
		+\frac{D^2}{3 f^2}
		I_{f+g}^{\mu,\Lambda \pi}
		%%%%%%%%%%%%
		+\frac{(D-F)^2}{2 f^2}
		I_{f+g}^{\mu,N K} , \\
		%%%%%%%%%++++++++++
		%%%%%%%%%++++++++++
		\Gamma_{f+g}^\mu(\Sigma^0)&{}=
		\frac{(D-F)^2}{4 f^2}
		I_{f+g}^{\mu,N K} 
		%%%%%%%%%%%%
		-\frac{(D+F)^2}{4 f^2}
		I_{f+g}^{\mu,\Xi K}, \\
		%%%%%%%%%++++++++++
		%%%%%%%%%++++++++++
		\Gamma_{f+g}^\mu(\Sigma^+)&{}=
		-\frac{F^2}{f^2}
		I_{f+g}^{\mu,\Sigma \pi} 
		%%%%%%%%%%%%
		-\frac{D^2}{3 f^2}
		I_{f+g}^{\mu,\Lambda \pi}
		%%%%%%%%%%%%
		-\frac{(D+F)^2}{2 f^2}
		I_{f+g}^{\mu,\Xi K}, 
	\end{align} 
	where
	\begin{equation}
		\begin{aligned}
			I_{f+g}^{\mu,BM} =
			\bar{u}(p^\prime)\tilde F(q)\int \frac{\dif^4 k}{(2\pi)^4}
			\frac{\tilde F(k)}{D_M(k)}
		  	\Big\lbrace {}&{}
			\frac{(2k-q)^\mu}{2kq-q^2} [\tilde F(k-q)-\tilde F(k)]
			\slashed{k}\gamma_5\frac{1}{\slashed{p'} -\slashed{k}-m_B} 
			 (-\slashed{k}+\slashed{q})\gamma_5
			\\ + {}&{} 
			\frac{(2k+q)^\mu}{2kq+q^2} [\tilde F(k+q)-\tilde F(k)]
			(\slashed{k}+\slashed{q})\gamma_5
			 \frac{1}{\slashed{p}-\slashed{k}-m_B}
			\slashed{k}\gamma_5 
			\Big\rbrace u(p).
		\end{aligned}
	\end{equation}

Now we show the expressions of one loop integrals for decuplet intermediate states.
The contribution for Fig.~\ref{fig.loop.diagrams}h can be written as
\begin{align}
	%%%+++++++++++++++++++
	\Gamma_{h}^\mu(\Sigma^-)={}&
		%%%+++++++++++++++++++
	\frac{\mathcal{C}^2}{12 f^2}
	I_{h}^{\mu,\Sigma^\ast \pi}
		%%%%%%%%%%%
	+\frac{\mathcal{C}^2}{6 f^2}
	%%%%%%%%%%%
	I_{h}^{\mu,\Delta K} ,\\
	%%%%%%%%%%%%%%%%%%
%%%+++++++++++++++++++
\Gamma_{h}^\mu(\Sigma^0)={}&
-\frac{\mathcal{C}^2}{12 f^2}
%%%%%%%%%%%
I_{h}^{\mu,\Sigma^\ast \pi}
%%++++++++++++++++
+
\frac{\mathcal{C}^2}{3 f^2}
I_{h}^{\mu,\Delta K}, \\
%%%+++++++++++++++++++
\Gamma_{h}^\mu(\Sigma^+)={}&
%%%+++++++++++++++++++
-\frac{\mathcal{C}^2}{12 f^2}
%%%%%%%%%%%
I_{h}^{\mu,\Sigma^\ast \pi}
%%%+++++++++++++++++++
-\frac{\mathcal{C}^2}{6 f^2} 
%%%%%%%%%%%
I_{h}^{\mu,\Xi^\ast \pi}
%%+++++++++
+
\frac{\mathcal{C}^2}{2 f^2}
%%%%%%%%%%%
I_{h}^{\mu,\Delta K},
\end{align}
where the integral $I_h^{\mu,TM}$ is expressed as
\begin{eqnarray}
I_h^{\mu,TM}&=&
\bar{u}(p^\prime) \tilde{F}(q)\int \frac{\dif^4 k}{(2 \pi)^4}
\frac{\tilde{F}(q+k)\tilde{F}(k)}{D_M(k)}
\frac{2(k+q)^\mu}{D_M (q+k)}
\\ && \times{} 
((k+q)^\sigma+z(\slashed{k}+\slashed{q})\gamma^\sigma)
\frac{1}{\slashed{p}-\slashed{k}-m_T}
S_{\sigma\rho}(p-k)
(-k^\rho-z\gamma^\rho\slashed{k})
u(p).
\end{eqnarray}
$m_T$ is the mass of the decuplet intermediate state and $S_{\sigma\rho}(p)$ is expressed as
\begin{equation}
S_{\sigma\rho}(p)=-g_{\sigma\rho}+
\frac{\gamma_\sigma\gamma_\rho}{3}+
\frac{p_\sigma p_\rho}{3{m_T}^2}+
\frac{\gamma_\sigma p_\rho-\gamma_\rho p_\sigma}{3 m_T}.
\end{equation}
The contribution for Fig.~\ref{fig.loop.diagrams}i is written as
\begin{align}
\Gamma_{i}^\mu(\Sigma^-)= {}& 
-\frac{\mathcal{C}^2 \left(\left(c_1+3 c_2+3\right) Q^2+
12 m_{\Sigma^\ast}^2\right)}{36 f^2 \left(4 m_{\Sigma^\ast}^2+Q^2\right)}
%%%%%%%%%%%
I_{i}^{\mu,\Sigma^\ast \pi}
%%%%%%%%%%%
-\frac{\mathcal{C}^2 \left(\left(c_1+3 c_2+3\right) Q^2+
12 m_{\Delta }^2\right)}{6 f^2 \left(4 m_{\Delta }^2+Q^2\right)}
%%%%%%%%%%%
I_{i}^{\mu,\Delta K}  \\ \nonumber
%%%+++++++++++++++++++
{}&{} 
-\frac{\mathcal{C}^2 \left(\left(c_1+3 c_2+3\right) Q^2+12 m_{\Xi^\ast}^2\right)}{18 f^2 \left(4 m_{\Xi^\ast}^2+Q^2\right)}
%%%+++++++++++++++++++
I_{i}^{\mu,\Xi^\ast K}, \\
%%%+++++++++++++++++++
\Gamma_{i}^\mu(\Sigma^0)= {}&{}
\frac{\mathcal{C}^2 \left(\left(c_1+3 c_2+3\right) Q^2+
	12 m_{\Delta }^2\right)}{9 f^2 \left(4 m_{\Delta }^2+Q^2\right)}
%%%%%%%%%%%
I_{i}^{\mu,\Delta K} 
%%%+++++++++++++++++++
-\frac{\mathcal{C}^2 \left(\left(c_1+3 c_2+3\right) Q^2+
	12 m_{\Xi^\ast}^2\right)}{36 f^2 \left(4 m_{\Xi^\ast}^2+Q^2\right)}
%%%%%%%%%%%
I_{i}^{\mu,\Xi^\ast K}, \\
%%%+++++++++++++++++++
\Gamma_{i}^\mu(\Sigma^+)={}&{}
\frac{\mathcal{C}^2 \left(\left(c_1+3 c_2+3\right) Q^2+12 m_{\Sigma^\ast}^2\right)}{36 f^2 \left(4 m_{\Sigma^\ast}^2+Q^2\right)}
%%%%%%%%%%%
I_{i}^{\mu,\Sigma^\ast \pi}
+\frac{7 \mathcal{C}^2 \left(\left(c_1+3 c_2+3\right) Q^2+12 m_{\Delta }^2\right)}{18 f^2 \left(4 m_{\Delta }^2+Q^2\right)}
%%%%%%%%%%%
I_{i}^{\mu,\Delta K},
%%%+++++++++++++++++++
\end{align}
where the integral $I_i^{\mu,TM}$ is written as
\begin{eqnarray}
I_i^{\mu,TM}&=&
\bar{u}(p^\prime) \tilde{F}(q)\int \frac{\dif^4 k}{(2\pi)^4}
\frac{\tilde{F}(k)^2}{D_M(k)}
(k^\sigma+z\slashed{k}\gamma^\sigma)
\\ && \times{}
\frac{1}{\slashed{p}^\prime-\slashed{k}-m_T}
S_{\sigma\alpha}(p^\prime-k)
\gamma^{\alpha\beta\mu}
\frac{1}{\slashed{p}-\slashed{k}-m_T}
S_{\beta\rho}(p-k)
(k^\rho+z\gamma^\rho\slashed{k})
u(p).
\end{eqnarray}
The contribution for Fig.~\ref{fig.loop.diagrams}j is written as
\begin{align}
%%%+++++++++++++++++++
\Gamma_{j}^\mu(\Sigma^-)=
{}&
\frac{ \left(c_1+3 c_2\right) \mathcal{C}^2 m_{\Sigma^\ast}}
{18 f^2 \left(4 m_{\Sigma^\ast}^2+Q^2\right)}
%%%%%%%%%%%
I_{j}^{\mu,\Sigma^\ast \pi}
%%%+++++++++++++++++++
+
\frac{ \left(c_1+3 c_2\right) \mathcal{C}^2 m_{\Delta }}
{3 f^2 \left(4 m_{\Delta }^2+Q^2\right)}
%%%%%%%%%%%
I_{j}^{\mu,\Delta K}
%%%%%%%%%%%%%%%%%%
+
\frac{ \left(c_1+3 c_2\right) \mathcal{C}^2 m_{\Xi^\ast}}{9 f^2 \left(4 m_{\Xi^\ast}^2+Q^2\right)}
%%%%%%%%%%%
I_{j}^{\mu,\Xi^\ast K}, \\
%%%+++++++++++++++++++
\Gamma_{j}^\mu(\Sigma^0)=
{}&
-\frac{2  \left(c_1+3 c_2\right) \mathcal{C}^2 m_{\Delta }}{9 f^2 \left(4 m_{\Delta }^2+Q^2\right)}
%%%%%%%%%%%
I_{j}^{\mu,\Delta K} 
%%%+++++++++++++++++++
+
\frac{ \left(c_1+3 c_2\right) \mathcal{C}^2 m_{\Xi^\ast}}{18 f^2 \left(4 m_{\Xi^\ast}^2+Q^2\right)}
%%%%%%%%%%%
I_{j}^{\mu,\Xi^\ast K}, \\
%%%+++++++++++++++++++
\Gamma_{j}^\mu(\Sigma^+)=
{}&
-\frac{ \left(c_1+3 c_2\right) \mathcal{C}^2 m_{\Sigma^\ast}}
{18 f^2 \left(4 m_{\Sigma^\ast}^2+Q^2\right)}
%%%%%%%%%%%
I_{j}^{\mu,\Sigma^\ast \pi}
%%%{}&
-\frac{7  \left(c_1+3 c_2\right) \mathcal{C}^2 m_{\Delta }}
{9 f^2 \left(4 m_{\Delta }^2+Q^2\right)}
%%%%%%%%%%%
I_{j}^{\mu,\Delta K},
%%%%%%%%%%%%%%%%%%
\end{align}
where the integral $I_j^{\mu,T M}$ is expressed as
\begin{eqnarray}
I_j^{\mu,T M}&=&
\bar{u}(p^\prime) \tilde{F}(q)\int \frac{\dif^4 k}{(2\pi)^4}
\frac{\tilde{F}(k)^2}{D_M(k)}
(k^\sigma+z\slashed{k}\gamma^\sigma)
\\ && \times{} 
\frac{1}{\slashed{p}^\prime-\slashed{k}-m_T}
S_{\sigma\nu}(p^\prime-k)
i \sigma^{\mu\lambda} q_\lambda
\frac{1}{\slashed{p}-\slashed{k}-m_T}
S^{\nu\rho}(p-k)
(k_\rho+z\gamma_\rho\slashed{k})
u(p).
\end{eqnarray}
The contribution for the intermediate octet-decuplet transition diagrams Figs.~\ref{fig.loop.diagrams}k and \ref{fig.loop.diagrams}l is expressed as
\begin{align} 
\Gamma_{k+l}^\mu (\Sigma^-)={}&
-\frac{c_1 \mathcal{C} F}{12 f^2 m_{\Sigma }}
%vvvvvvvvvvvvvvvvvvv
I_{k+l}^{\mu,\Sigma^\ast \Sigma \pi}
%vvvvvvvvvvvvvvvvvvv
+
\frac{c_1 \mathcal{C} D}{12 f^2 m_{\Lambda }}
I_{k+l}^{\mu,\Sigma^\ast \Lambda \pi} 
%%%%%%%%%%%%%%%%%%%%%%%
+
\frac{c_1 \mathcal{C} (D-F)}{6 f^2 m_N}
%vvvvvvvvvvvvvvvvvvv
I_{k+l}^{\mu,\Delta N K}, \\
%%%%%%%%%%%%%%%%%%%%%%%
\Gamma_{k+l}^\mu (\Sigma^0)={}&
-\frac{c_1 \mathcal{C} F}{6 f^2 m_{\Sigma }}
%vvvvvvvvvvvvvvvvvvv
I_{k+l}^{\mu,\Sigma^\ast \Sigma \pi}
%%%%%%%%%%%%%%%%%%%%%%
-\frac{c_1 \mathcal{C} (D+F)}{12 f^2 m_{\Xi }}
%vvvvvvvvvvvvvvvvvvv
I_{k+l}^{\mu,\Xi^\ast \Xi K} , \\
%%%%%%%%%%+++++++++++++++++
\Gamma_{k+l}^\mu (\Sigma^+)={}&
%%%%%%%
-\frac{c_1 \mathcal{C} F}{4 f^2 m_{\Sigma }}
%vvvvvvvvvvvvvvvvvvv
I_{k+l}^{\mu,\Sigma^\ast \Sigma \pi} 
%%%%%%%%%%%%%%%%%%%%%%
-\frac{c_1 \mathcal{C} D}{12 f^2 m_{\Lambda }}
%vvvvvvvvvvvvvvvvvvv
I_{k+l}^{\mu,\Sigma^\ast \Lambda \pi}
%%%%%%%%%%%%%%%%%%%%%%
-\frac{c_1 \mathcal{C} (D-F)}{6 f^2 m_N}
%vvvvvvvvvvvvvvvvvvv
I_{k+l}^{\mu,\Delta N K}
%%%%%%%%%%%
-\frac{c_1 \mathcal{C} (D+F)}{6 f^2 m_{\Xi }}
%vvvvvvvvvvvvvvvvvvv
I_{k+l}^{\mu,\Xi^\ast \Xi K} ,
\end{align}
where the integral $I_{k+l}^{\mu,TBM}$ is written as
\begin{equation}
 \begin{aligned}
I_{k+l}^{\mu,TBM} = {}&
\bar{u}(p^\prime) 
\tilde{F}(q)
\int \frac{\dif^4 k}{(2\pi)^4}
\frac{\tilde{F}(k)^2} {D_M(k)} \Big\lbrace 
{}
\slashed{k}\gamma_5
\frac{1}{\slashed{p}^\prime-\slashed{k}-m_B}
(-\slashed{q})\gamma_5
\frac{1}{\slashed{p}-\slashed{k}-m_T}
S^{\mu\rho}(p-k)
(k_\rho+z\gamma_\rho\slashed{k}) 
\\ +{} &
\slashed{k}\gamma_5
\frac{1}{\slashed{p}^\prime-\slashed{k}-m_B}
\gamma^\mu\gamma_5 q_\nu
\frac{1}{\slashed{p}-\slashed{k}-m_T}
S^{\nu\rho}(p-k)
(k_\rho+z\gamma_\rho\slashed{k})
\\ +{} & 
(k_\nu+z\slashed{k}\gamma_\nu)
\frac{1}{\slashed{p}^\prime-\slashed{k}-m_T}
S^{\nu\rho}(p^\prime-k)
(-q_\rho) \gamma^\mu\gamma_5
\frac{1}{\slashed{p}-\slashed{k}-m_B}
\slashed{k}\gamma_5 
\\ +{} &
(k_\nu+z\slashed{k}\gamma_\nu)
\frac{1}{\slashed{p}^\prime-\slashed{k}-m_T}
S^{\nu\mu}(p^\prime-k)
\slashed{q}\gamma_5
\frac{1}{\slashed{p}-\slashed{k}-m_B}
\slashed{k}\gamma_5
\Big\rbrace 
u(p).
\end{aligned}
\end{equation}
The contribution for the Kroll-Ruderman diagrams Figs.~\ref{fig.loop.diagrams}m and \ref{fig.loop.diagrams}n is written as
\begin{align} 
\Gamma_{m+n}^\mu(\Sigma^-)={}&
\frac{\mathcal{C}^2}{12 f^2}
%vvvvvvvvvvvvvvvvvvv
I_{m+n}^{\mu,\Sigma^\ast \pi}
%vvvvvvvvvvvvvvvvvvv
+
\frac{\mathcal{C}^2}{6 f^2}
%vvvvvvvvvvvvvvvvvvv
I_{m+n}^{\mu,\Delta K} ,\\
%%%%%%%%%%%%%%%%%%%%%%%%%%%%%%%%%%%%%%%%%
\Gamma_{m+n}^\mu(\Sigma^0)={}&
\frac{\mathcal{C}^2}{3 f^2}
%vvvvvvvvvvvvvvvvvvv
I_{m+n}^{\mu,\Delta K}
%vvvvvvvvvvvvvvvvvvv
-\frac{\mathcal{C}^2}{12 f^2}
%vvvvvvvvvvvvvvvvvvv
I_{m+n}^{\mu,\Sigma^\ast K} ,\\
%%%%%%%%%%%%%%%%%%%%%%%%%%%%%%%%%%%%%%%%%
\Gamma_{m+n}^\mu(\Sigma^+)={}&
%vvvvvvvvvvvvvvvvvvv
-\frac{\mathcal{C}^2}{12 f^2}
%vvvvvvvvvvvvvvvvvvv
I_{m+n}^{\mu,\Sigma^\ast \pi} 
%vvvvvvvvvvvvvvvvvvv
+
\frac{\mathcal{C}^2}{2 f^2}
%vvvvvvvvvvvvvvvvvvv
I_{m+n}^{\mu,\Delta K}
%%+++++++++++++
-\frac{\mathcal{C}^2}{6 f^2}
I_{m+n}^{\mu,\Xi^\ast K} ,
\end{align}
where the integral $I_{m+n}^{\mu,TM}$ is written as
\begin{equation}
\begin{aligned}
I_{m+n}^{\mu,TM}=
\bar{u}(p^\prime)
\tilde{F}(q)
\int \frac{\dif^4 k}{(2\pi)^4}
\frac{\tilde{F}(k)^2}{D_M(k)}
\Big\lbrace
&(k_\sigma+z\slashed{k}\gamma_\sigma)
\frac{1}{\slashed{p}^\prime-\slashed{k}-m_T}
S^{\sigma \rho}(p^\prime-k)
\left({g_{\rho}}^{\mu}+z\gamma_\rho \gamma^\mu\right)+{} 
\\ &
({g^{\mu}}_{\sigma}+z\gamma^\mu\gamma_\sigma)
\frac{1}{\slashed{p}-\slashed{k}-m_T}
S^{\sigma \rho}(p-k)
\left(k_\rho+z\gamma_\rho\slashed{k}\right)
\Big \rbrace 
u(p).
\end{aligned}
\end{equation}
The contribution for the additional diagrams with intermediate decuplet states Figs.~\ref{fig.loop.diagrams}o and \ref{fig.loop.diagrams}p is expressed as
\begin{align} 
\Gamma_{o+p}^\mu(\Sigma^-)={}&
%%%++++++++++++
\frac{\mathcal{C}^2}{12 f^2}
%vvvvvvvvvvvvvvvvvvv
I_{o+p}^{\mu,\Sigma^\ast \pi}
%vvvvvvvvvvvvvvvvvvv
+
\frac{\mathcal{C}^2}{6 f^2}
%vvvvvvvvvvvvvvvvvvv
I_{o+p}^{\mu,\Delta K} ,\\
%%%%%%%%%%%%%%%%%%%%%%%%%%%%%%%%%%%%%%%%%
\Gamma_{o+p}^\mu(\Sigma^0)={}&
\frac{\mathcal{C}^2}{3 f^2}
%vvvvvvvvvvvvvvvvvvv
I_{o+p}^{\mu,\Delta K}
%vvvvvvvvvvvvvvvvvvv
-\frac{\mathcal{C}^2}{12 f^2}
%vvvvvvvvvvvvvvvvvvv
I_{o+p}^{\mu,\Sigma^\ast K} , \\
%%%%%%%%%%%%%%%%%%%%%%%%%%%%%%%%%%%%%%%%%
\Gamma_{o+p}^\mu(\Sigma^+)={}&
%vvvvvvvvvvvvvvvvvvv
-\frac{\mathcal{C}^2}{12 f^2}
%vvvvvvvvvvvvvvvvvvv
I_{o+p}^{\mu,\Sigma^\ast \pi} 
%%+++++++++++++
+
\frac{\mathcal{C}^2}{2 f^2}
%vvvvvvvvvvvvvvvvvvv
I_{o+p}^{\mu,\Delta K}
%vvvvvvvvvvvvvvvvvvv
-\frac{\mathcal{C}^2}{6 f^2}
%vvvvvvvvvvvvvvvvvvv
I_{o+p}^{\mu,\Xi^\ast K} ,
\end{align}
where the integral $I_{o+p}^{\mu,TM}$ is written as
\begin{equation}
\begin{aligned}
I_{o+p}^{\mu,TM}=
{}&
\bar{u}(p^\prime)
\tilde{F}(q)
\int \frac{\dif^4 k}{(2\pi)^4}
\frac{\tilde{F}(k)}{D_M(k)}
\\ &
\Big\lbrace 
\frac{(-2k+q)^\mu}{-2kq+q^2}
\left(\tilde{F}(k-q)-\tilde{F}(k)\right)
(k_\sigma+z\slashed{k}\gamma_\sigma)
\frac{1}{\slashed{p}^\prime-\slashed{k}-m_T}
S^{\sigma\rho}(p^\prime-k)
% \times{}\\ &
\left((k-q)_\rho+z\gamma_\rho(\slashed{k}-\slashed{q})\right)
%%%%%%%%%%%%part2%%%%%%%%%%%%%%%%%%
\\ & +{}  
\frac{(2k+q)^\mu}{2kq+q^2}
\left(\tilde{F}(k+q)-\tilde{F}(k)\right)
\left((k+q)_\sigma+
z(\slashed{k}+\slashed{q})\gamma_\sigma\right)
\frac{1}{\slashed{p}-\slashed{k}-m_T}
S^{\sigma\rho}(p-k)
\left(k_\rho+z\gamma_\rho\slashed{k}\right) 
%\\&
\Big \rbrace
u(p).
\end{aligned}
\end{equation}
Using Package-X \cite{tool.packagex} to simplify the loop integral, we can get the results for the Dirac and Pauli form factors.

%%%++++++++++++++++++
%\bibliography{octet.ff.2}

\begin{thebibliography}{74}%
	\makeatletter
	\providecommand \@ifxundefined [1]{%
		\@ifx{#1\undefined}
	}%
	\providecommand \@ifnum [1]{%
		\ifnum #1\expandafter \@firstoftwo
		\else \expandafter \@secondoftwo
		\fi
	}%
	\providecommand \@ifx [1]{%
		\ifx #1\expandafter \@firstoftwo
		\else \expandafter \@secondoftwo
		\fi
	}%
	\providecommand \natexlab [1]{#1}%
	\providecommand \enquote  [1]{``#1''}%
	\providecommand \bibnamefont  [1]{#1}%
	\providecommand \bibfnamefont [1]{#1}%
	\providecommand \citenamefont [1]{#1}%
	\providecommand \href@noop [0]{\@secondoftwo}%
	\providecommand \href [0]{\begingroup \@sanitize@url \@href}%
	\providecommand \@href[1]{\@@startlink{#1}\@@href}%
	\providecommand \@@href[1]{\endgroup#1\@@endlink}%
	\providecommand \@sanitize@url [0]{\catcode `\\12\catcode `\$12\catcode
		`\&12\catcode `\#12\catcode `\^12\catcode `\_12\catcode `\%12\relax}%
	\providecommand \@@startlink[1]{}%
	\providecommand \@@endlink[0]{}%
	\providecommand \url  [0]{\begingroup\@sanitize@url \@url }%
	\providecommand \@url [1]{\endgroup\@href {#1}{\urlprefix }}%
	\providecommand \urlprefix  [0]{URL }%
	\providecommand \Eprint [0]{\href }%
	\providecommand \doibase [0]{https://doi.org/}%
	\providecommand \selectlanguage [0]{\@gobble}%
	\providecommand \bibinfo  [0]{\@secondoftwo}%
	\providecommand \bibfield  [0]{\@secondoftwo}%
	\providecommand \translation [1]{[#1]}%
	\providecommand \BibitemOpen [0]{}%
	\providecommand \bibitemStop [0]{}%
	\providecommand \bibitemNoStop [0]{.\EOS\space}%
	\providecommand \EOS [0]{\spacefactor3000\relax}%
	\providecommand \BibitemShut  [1]{\csname bibitem#1\endcsname}%
	\let\auto@bib@innerbib\@empty
	%</preamble>
	\bibitem [{\citenamefont {Camsonne}\ \emph {et~al.}(2014)\citenamefont
		{Camsonne} \emph {et~al.}}]{Camsonne}%
	\BibitemOpen
	\bibfield  {author} {\bibinfo {author} {\bibfnamefont {A.}~\bibnamefont
			{Camsonne}} \emph {et~al.} (\bibinfo {collaboration} {Jefferson Lab Hall
			A}),\ }\bibfield  {title} {\bibinfo {title} {{JLab Measurement of the $^4$He
				Charge Form Factor at Large Momentum Transfers}},\ }\href
	{https://doi.org/10.1103/PhysRevLett.112.132503} {\bibfield  {journal}
		{\bibinfo  {journal} {Phys. Rev. Lett.}\ }\textbf {\bibinfo {volume} {112}},\
		\bibinfo {pages} {132503} (\bibinfo {year} {2014})},\ \Eprint
	{https://arxiv.org/abs/1309.5297} {arXiv:1309.5297 [nucl-ex]} \BibitemShut
	{NoStop}%
	\bibitem [{\citenamefont {Sirunyan}\ \emph {et~al.}(2017)\citenamefont
		{Sirunyan} \emph {et~al.}}]{CMS}%
	\BibitemOpen
	\bibfield  {author} {\bibinfo {author} {\bibfnamefont {A.~M.}\ \bibnamefont
			{Sirunyan}} \emph {et~al.} (\bibinfo {collaboration} {CMS}),\ }\bibfield
	{title} {\bibinfo {title} {{Measurement of the triple-differential dijet
				cross section in proton-proton collisions at $\sqrt{s}=8\,\text {TeV} $ and
				constraints on parton distribution functions}},\ }\href
	{https://doi.org/10.1140/epjc/s10052-017-5286-7} {\bibfield  {journal}
		{\bibinfo  {journal} {Eur. Phys. J. C}\ }\textbf {\bibinfo {volume} {77}},\
		\bibinfo {pages} {746} (\bibinfo {year} {2017})},\ \Eprint
	{https://arxiv.org/abs/1705.02628} {arXiv:1705.02628 [hep-ex]} \BibitemShut
	{NoStop}%
	\bibitem [{\citenamefont {Xiong}\ \emph {et~al.}(2019)\citenamefont {Xiong}
		\emph {et~al.}}]{Xiong}%
	\BibitemOpen
	\bibfield  {author} {\bibinfo {author} {\bibfnamefont {W.}~\bibnamefont
			{Xiong}} \emph {et~al.},\ }\bibfield  {title} {\bibinfo {title} {{A small
				proton charge radius from an electron--proton scattering experiment}},\
	}\href {https://doi.org/10.1038/s41586-019-1721-2} {\bibfield  {journal}
		{\bibinfo  {journal} {Nature}\ }\textbf {\bibinfo {volume} {575}},\ \bibinfo
		{pages} {147} (\bibinfo {year} {2019})}\BibitemShut {NoStop}%
	\bibitem [{\citenamefont {Bernauer}\ \emph {et~al.}(2010)\citenamefont
		{Bernauer} \emph {et~al.}}]{Bernauer}%
	\BibitemOpen
	\bibfield  {author} {\bibinfo {author} {\bibfnamefont {J.}~\bibnamefont
			{Bernauer}} \emph {et~al.} (\bibinfo {collaboration} {A1}),\ }\bibfield
	{title} {\bibinfo {title} {{High-precision determination of the electric and
				magnetic form factors of the proton}},\ }\href
	{https://doi.org/10.1103/PhysRevLett.105.242001} {\bibfield  {journal}
		{\bibinfo  {journal} {Phys. Rev. Lett.}\ }\textbf {\bibinfo {volume} {105}},\
		\bibinfo {pages} {242001} (\bibinfo {year} {2010})},\ \Eprint
	{https://arxiv.org/abs/1007.5076} {arXiv:1007.5076 [nucl-ex]} \BibitemShut
	{NoStop}%
	\bibitem [{\citenamefont {Kubodera}\ \emph {et~al.}(1985)\citenamefont
		{Kubodera}, \citenamefont {Kohyama}, \citenamefont {Oikawa} \emph
		{et~al.}}]{theory.bag-model}%
	\BibitemOpen
	\bibfield  {author} {\bibinfo {author} {\bibfnamefont {K.}~\bibnamefont
			{Kubodera}}, \bibinfo {author} {\bibfnamefont {Y.}~\bibnamefont {Kohyama}},
		\bibinfo {author} {\bibfnamefont {K.}~\bibnamefont {Oikawa}}, \emph
		{et~al.},\ }\bibfield  {title} {\bibinfo {title} {Weak-interaction form
			factors of octet baryons in the cloudy bag model},\ }\href
	{https://doi.org/https://doi.org/10.1016/0375-9474(85)90334-3} {\bibfield
		{journal} {\bibinfo  {journal} {Nucl. Phys. A}\ }\textbf {\bibinfo {volume}
			{439}},\ \bibinfo {pages} {695} (\bibinfo {year} {1985})}\BibitemShut
	{NoStop}%
	\bibitem [{\citenamefont {Dahiya}\ \emph {et~al.}(2009)\citenamefont {Dahiya},
		\citenamefont {Sharma}, \citenamefont {Chatley},\ and\ \citenamefont
		{Manmoha}}]{theory.Chiral-Constituent-Quark-Model}%
	\BibitemOpen
	\bibfield  {author} {\bibinfo {author} {\bibfnamefont {H.}~\bibnamefont
			{Dahiya}}, \bibinfo {author} {\bibfnamefont {N.}~\bibnamefont {Sharma}},
		\bibinfo {author} {\bibfnamefont {P.~K.}\ \bibnamefont {Chatley}},\ and\
		\bibinfo {author} {\bibfnamefont {G.}~\bibnamefont {Manmoha}},\ }\bibfield
	{title} {\bibinfo {title} {Semi‐leptonic octet baryon weak axial‐vector
			form factors in the chiral constituent quark model},\ }\href
	{https://doi.org/10.1063/1.3215665} {\bibfield  {journal} {\bibinfo
			{journal} {AIP Conf. Proc.}\ }\textbf {\bibinfo {volume} {1149}},\ \bibinfo
		{pages} {361} (\bibinfo {year} {2009})}\BibitemShut {NoStop}%
	\bibitem [{\citenamefont {Kim}\ and\ \citenamefont {Kim}(2018)}]{theory.1diNc}%
	\BibitemOpen
	\bibfield  {author} {\bibinfo {author} {\bibfnamefont {J.-Y.}\ \bibnamefont
			{Kim}}\ and\ \bibinfo {author} {\bibfnamefont {H.-C.}\ \bibnamefont {Kim}},\
	}\bibfield  {title} {\bibinfo {title} {{Electromagnetic form factors of
				singly heavy baryons in the self-consistent SU(3) chiral quark-soliton
				model}},\ }\href {https://doi.org/10.1103/PhysRevD.97.114009} {\bibfield
		{journal} {\bibinfo  {journal} {Phys. Rev. D}\ }\textbf {\bibinfo {volume}
			{97}},\ \bibinfo {pages} {114009} (\bibinfo {year} {2018})}\BibitemShut
	{NoStop}%
	\bibitem [{\citenamefont {Ito}\ \emph {et~al.}(2009)\citenamefont {Ito},
		\citenamefont {Bentz}, \citenamefont {Clo{\"e}t}, \citenamefont {Thomas},\
		and\ \citenamefont {Yazaki}}]{Ito}%
	\BibitemOpen
	\bibfield  {author} {\bibinfo {author} {\bibfnamefont {T.}~\bibnamefont
			{Ito}}, \bibinfo {author} {\bibfnamefont {W.}~\bibnamefont {Bentz}}, \bibinfo
		{author} {\bibfnamefont {I.}~\bibnamefont {Clo{\"e}t}}, \bibinfo {author}
		{\bibfnamefont {A.}~\bibnamefont {Thomas}},\ and\ \bibinfo {author}
		{\bibfnamefont {K.}~\bibnamefont {Yazaki}},\ }\bibfield  {title} {\bibinfo
		{title} {{The NJL-jet model for quark fragmentation functions}},\ }\href
	{https://doi.org/10.1103/PhysRevD.80.074008} {\bibfield  {journal} {\bibinfo
			{journal} {Phys. Rev. D}\ }\textbf {\bibinfo {volume} {80}},\ \bibinfo
		{pages} {074008} (\bibinfo {year} {2009})},\ \Eprint
	{https://arxiv.org/abs/0906.5362} {arXiv:0906.5362 [nucl-th]} \BibitemShut
	{NoStop}%
	\bibitem [{\citenamefont {Ohlsson}\ and\ \citenamefont
		{Snellman}(1999)}]{theory.chiral-quark.model}%
	\BibitemOpen
	\bibfield  {author} {\bibinfo {author} {\bibfnamefont {T.}~\bibnamefont
			{Ohlsson}}\ and\ \bibinfo {author} {\bibfnamefont {H.}~\bibnamefont
			{Snellman}},\ }\bibfield  {title} {\bibinfo {title} {Weak form factors for
			semileptonic octet baryon decays in the chiral quark model},\ }\href
	{https://doi.org/10.1007/s100529800908} {\bibfield  {journal} {\bibinfo
			{journal} {Eur. Phys. J. C}\ }\textbf {\bibinfo {volume} {6}},\ \bibinfo
		{pages} {285} (\bibinfo {year} {1999})}\BibitemShut {NoStop}%
	\bibitem [{\citenamefont {Yang}\ \emph {et~al.}(2019)\citenamefont {Yang},
		\citenamefont {Chen},\ and\ \citenamefont
		{Lu}}]{theory.extended-vector-meson}%
	\BibitemOpen
	\bibfield  {author} {\bibinfo {author} {\bibfnamefont {Y.}~\bibnamefont
			{Yang}}, \bibinfo {author} {\bibfnamefont {D.-Y.}\ \bibnamefont {Chen}},\
		and\ \bibinfo {author} {\bibfnamefont {Z.}~\bibnamefont {Lu}},\ }\bibfield
	{title} {\bibinfo {title} {{The electromagnetic form factors of $\Lambda$
				hyperon in the vector meson dominance model}},\ }\bibfield  {journal}
	{\bibinfo  {journal} {Phys. Rev. D}\ }\textbf {\bibinfo {volume} {100}},\
	\href {https://doi.org/10.1103/PhysRevD.100.073007}
	{10.1103/PhysRevD.100.073007} (\bibinfo {year} {2019}),\ \Eprint
	{https://arxiv.org/abs/1902.01242} {arXiv:1902.01242 [hep-ph]} \BibitemShut
	{NoStop}%
	\bibitem [{\citenamefont {An}\ and\ \citenamefont
		{Saghai}(2013)}]{theory.extended-chiral-quark-model}%
	\BibitemOpen
	\bibfield  {author} {\bibinfo {author} {\bibfnamefont {C.~S.}\ \bibnamefont
			{An}}\ and\ \bibinfo {author} {\bibfnamefont {B.}~\bibnamefont {Saghai}},\
	}\bibfield  {title} {\bibinfo {title} {Strangeness magnetic form factor of
			the proton in the extended chiral quark model},\ }\href
	{https://doi.org/10.1103/PhysRevC.88.025206} {\bibfield  {journal} {\bibinfo
			{journal} {Phys. Rev. C}\ }\textbf {\bibinfo {volume} {88}},\ \bibinfo
		{pages} {025206} (\bibinfo {year} {2013})}\BibitemShut {NoStop}%
	\bibitem [{\citenamefont {Faustov}\ and\ \citenamefont
		{Galkin}(2018)}]{theory.quark-diquark}%
	\BibitemOpen
	\bibfield  {author} {\bibinfo {author} {\bibfnamefont {R.~N.}\ \bibnamefont
			{Faustov}}\ and\ \bibinfo {author} {\bibfnamefont {V.~O.}\ \bibnamefont
			{Galkin}},\ }\bibfield  {title} {\bibinfo {title} {{Relativistic description
				of the ${\mathrm{\ensuremath{\Xi}}}_{b}$ baryon semileptonic decays}},\
	}\href {https://doi.org/10.1103/PhysRevD.98.093006} {\bibfield  {journal}
		{\bibinfo  {journal} {Phys. Rev. D}\ }\textbf {\bibinfo {volume} {98}},\
		\bibinfo {pages} {093006} (\bibinfo {year} {2018})}\BibitemShut {NoStop}%
	\bibitem [{\citenamefont {Fuchs}\ \emph {et~al.}(2004)\citenamefont {Fuchs},
		\citenamefont {Gegelia},\ and\ \citenamefont
		{Scherer}}]{theory.chpt-range.1}%
	\BibitemOpen
	\bibfield  {author} {\bibinfo {author} {\bibfnamefont {T.}~\bibnamefont
			{Fuchs}}, \bibinfo {author} {\bibfnamefont {J.}~\bibnamefont {Gegelia}},\
		and\ \bibinfo {author} {\bibfnamefont {S.}~\bibnamefont {Scherer}},\
	}\bibfield  {title} {\bibinfo {title} {Electromagnetic form factors of the
			nucleon in chiral perturbation theory},\ }\href
	{https://doi.org/10.1088/0954-3899/30/10/008} {\bibfield  {journal} {\bibinfo
			{journal} {J. Phys. G}\ }\textbf {\bibinfo {volume} {30}},\ \bibinfo {pages}
		{1407} (\bibinfo {year} {2004})}\BibitemShut {NoStop}%
	\bibitem [{\citenamefont {Kubis}\ and\ \citenamefont
		{Meissner}(2001{\natexlab{a}})}]{theory.chpt-range.2}%
	\BibitemOpen
	\bibfield  {author} {\bibinfo {author} {\bibfnamefont {B.}~\bibnamefont
			{Kubis}}\ and\ \bibinfo {author} {\bibfnamefont {U.-G.}\ \bibnamefont
			{Meissner}},\ }\bibfield  {title} {\bibinfo {title} {Low-energy analysis of
			the nucleon electromagnetic form factors},\ }\href
	{https://doi.org/https://doi.org/10.1016/S0375-9474(00)00378-X} {\bibfield
		{journal} {\bibinfo  {journal} {Nucl. Phys. A}\ }\textbf {\bibinfo {volume}
			{679}},\ \bibinfo {pages} {698} (\bibinfo {year}
		{2001}{\natexlab{a}})}\BibitemShut {NoStop}%
	\bibitem [{\citenamefont {Wang}\ \emph {et~al.}(2007)\citenamefont {Wang},
		\citenamefont {Leinweber}, \citenamefont {Thomas},\ and\ \citenamefont
		{Young}}]{theory.FRR.0}%
	\BibitemOpen
	\bibfield  {author} {\bibinfo {author} {\bibfnamefont {P.}~\bibnamefont
			{Wang}}, \bibinfo {author} {\bibfnamefont {D.~B.}\ \bibnamefont {Leinweber}},
		\bibinfo {author} {\bibfnamefont {A.~W.}\ \bibnamefont {Thomas}},\ and\
		\bibinfo {author} {\bibfnamefont {R.~D.}\ \bibnamefont {Young}},\ }\bibfield
	{title} {\bibinfo {title} {Chiral extrapolation of nucleon magnetic form
			factors},\ }\href {https://doi.org/10.1103/PhysRevD.75.073012} {\bibfield
		{journal} {\bibinfo  {journal} {Phys. Rev. D}\ }\textbf {\bibinfo {volume}
			{75}},\ \bibinfo {pages} {073012} (\bibinfo {year} {2007})}\BibitemShut
	{NoStop}%
	\bibitem [{\citenamefont {Wang}\ \emph
		{et~al.}(2009{\natexlab{a}})\citenamefont {Wang}, \citenamefont {Leinweber},
		\citenamefont {Thomas},\ and\ \citenamefont {Young}}]{theory.FRR.1}%
	\BibitemOpen
	\bibfield  {author} {\bibinfo {author} {\bibfnamefont {P.}~\bibnamefont
			{Wang}}, \bibinfo {author} {\bibfnamefont {D.~B.}\ \bibnamefont {Leinweber}},
		\bibinfo {author} {\bibfnamefont {A.~W.}\ \bibnamefont {Thomas}},\ and\
		\bibinfo {author} {\bibfnamefont {R.~D.}\ \bibnamefont {Young}},\ }\bibfield
	{title} {\bibinfo {title} {{Strange magnetic form factor of the proton at
				${Q}^{2}=0.23 {\mathrm{GeV}}^{2}$}},\ }\href
	{https://doi.org/10.1103/PhysRevC.79.065202} {\bibfield  {journal} {\bibinfo
			{journal} {Phys. Rev. C}\ }\textbf {\bibinfo {volume} {79}},\ \bibinfo
		{pages} {065202} (\bibinfo {year} {2009}{\natexlab{a}})}\BibitemShut
	{NoStop}%
	\bibitem [{\citenamefont {Wang}\ \emph {et~al.}(2012)\citenamefont {Wang},
		\citenamefont {Leinweber}, \citenamefont {Thomas},\ and\ \citenamefont
		{Young}}]{theory.FRR.2}%
	\BibitemOpen
	\bibfield  {author} {\bibinfo {author} {\bibfnamefont {P.}~\bibnamefont
			{Wang}}, \bibinfo {author} {\bibfnamefont {D.~B.}\ \bibnamefont {Leinweber}},
		\bibinfo {author} {\bibfnamefont {A.~W.}\ \bibnamefont {Thomas}},\ and\
		\bibinfo {author} {\bibfnamefont {R.~D.}\ \bibnamefont {Young}},\ }\bibfield
	{title} {\bibinfo {title} {Chiral extrapolation of nucleon magnetic moments
			at next-to-leading-order},\ }\href
	{https://doi.org/10.1103/PhysRevD.86.094038} {\bibfield  {journal} {\bibinfo
			{journal} {Phys. Rev. D}\ }\textbf {\bibinfo {volume} {86}},\ \bibinfo
		{pages} {094038} (\bibinfo {year} {2012})}\BibitemShut {NoStop}%
	\bibitem [{\citenamefont {Hall}\ \emph {et~al.}(2014)\citenamefont {Hall},
		\citenamefont {Leinweber},\ and\ \citenamefont {Young}}]{theory.FRR.3}%
	\BibitemOpen
	\bibfield  {author} {\bibinfo {author} {\bibfnamefont {J.~M.~M.}\
			\bibnamefont {Hall}}, \bibinfo {author} {\bibfnamefont {D.~B.}\ \bibnamefont
			{Leinweber}},\ and\ \bibinfo {author} {\bibfnamefont {R.}~\bibnamefont
			{Young}},\ }\bibfield  {title} {\bibinfo {title} {Finite-volume and partial
			quenching effects in the magnetic polarizability of the neutron},\ }\href
	{https://doi.org/10.1103/PhysRevD.89.054511} {\bibfield  {journal} {\bibinfo
			{journal} {Phys. Rev. D}\ }\textbf {\bibinfo {volume} {89}},\ \bibinfo
		{pages} {054511} (\bibinfo {year} {2014})}\BibitemShut {NoStop}%
	\bibitem [{\citenamefont {Wang}\ \emph {et~al.}(2014)\citenamefont {Wang},
		\citenamefont {Leinweber},\ and\ \citenamefont {Thomas}}]{theory.FRR.4}%
	\BibitemOpen
	\bibfield  {author} {\bibinfo {author} {\bibfnamefont {P.}~\bibnamefont
			{Wang}}, \bibinfo {author} {\bibfnamefont {D.~B.}\ \bibnamefont
			{Leinweber}},\ and\ \bibinfo {author} {\bibfnamefont {A.~W.}\ \bibnamefont
			{Thomas}},\ }\bibfield  {title} {\bibinfo {title} {Strange magnetic form
			factor of the nucleon in a chiral effective model at next to leading order},\
	}\href {https://doi.org/10.1103/PhysRevD.89.033008} {\bibfield  {journal}
		{\bibinfo  {journal} {Phys. Rev. D}\ }\textbf {\bibinfo {volume} {89}},\
		\bibinfo {pages} {033008} (\bibinfo {year} {2014})}\BibitemShut {NoStop}%
	\bibitem [{\citenamefont {Wang}\ \emph {et~al.}(2015)\citenamefont {Wang},
		\citenamefont {Leinweber},\ and\ \citenamefont {Thomas}}]{theory.FRR.5}%
	\BibitemOpen
	\bibfield  {author} {\bibinfo {author} {\bibfnamefont {P.}~\bibnamefont
			{Wang}}, \bibinfo {author} {\bibfnamefont {D.~B.}\ \bibnamefont
			{Leinweber}},\ and\ \bibinfo {author} {\bibfnamefont {A.~W.}\ \bibnamefont
			{Thomas}},\ }\bibfield  {title} {\bibinfo {title} {Pure sea-quark
			contributions to the magnetic form factors of $\mathrm{\ensuremath{\Sigma}}$
			baryons},\ }\href {https://doi.org/10.1103/PhysRevD.92.034508} {\bibfield
		{journal} {\bibinfo  {journal} {Phys. Rev. D}\ }\textbf {\bibinfo {volume}
			{92}},\ \bibinfo {pages} {034508} (\bibinfo {year} {2015})}\BibitemShut
	{NoStop}%
	\bibitem [{\citenamefont {Li}\ \emph {et~al.}(2016)\citenamefont {Li},
		\citenamefont {Wang}, \citenamefont {Leinweber},\ and\ \citenamefont
		{Thomas}}]{theory.FRR.6}%
	\BibitemOpen
	\bibfield  {author} {\bibinfo {author} {\bibfnamefont {H.}~\bibnamefont
			{Li}}, \bibinfo {author} {\bibfnamefont {P.}~\bibnamefont {Wang}}, \bibinfo
		{author} {\bibfnamefont {D.~B.}\ \bibnamefont {Leinweber}},\ and\ \bibinfo
		{author} {\bibfnamefont {A.~W.}\ \bibnamefont {Thomas}},\ }\bibfield  {title}
	{\bibinfo {title} {Spin of the proton in chiral effective field theory},\
	}\href {https://doi.org/10.1103/PhysRevC.93.045203} {\bibfield  {journal}
		{\bibinfo  {journal} {Phys. Rev. C}\ }\textbf {\bibinfo {volume} {93}},\
		\bibinfo {pages} {045203} (\bibinfo {year} {2016})}\BibitemShut {NoStop}%
	\bibitem [{\citenamefont {Li}\ and\ \citenamefont {Wang}(2016)}]{theory.FRR.7}%
	\BibitemOpen
	\bibfield  {author} {\bibinfo {author} {\bibfnamefont {H.}~\bibnamefont
			{Li}}\ and\ \bibinfo {author} {\bibfnamefont {P.}~\bibnamefont {Wang}},\
	}\bibfield  {title} {\bibinfo {title} {Chiral extrapolation of nucleon axial
			charge ga in effective field theory},\ }\href
	{https://doi.org/10.1088/1674-1137/40/12/123106} {\bibfield  {journal}
		{\bibinfo  {journal} {Chin. Phys. C}\ }\textbf {\bibinfo {volume} {40}},\
		\bibinfo {pages} {123106} (\bibinfo {year} {2016})}\BibitemShut {NoStop}%
	\bibitem [{\citenamefont {Allton}\ \emph {et~al.}(2005)\citenamefont {Allton},
		\citenamefont {Armour}, \citenamefont {Leinweber}, \citenamefont {Thomas},\
		and\ \citenamefont {Young}}]{theory.FRR.8}%
	\BibitemOpen
	\bibfield  {author} {\bibinfo {author} {\bibfnamefont {C.~R.}\ \bibnamefont
			{Allton}}, \bibinfo {author} {\bibfnamefont {W.}~\bibnamefont {Armour}},
		\bibinfo {author} {\bibfnamefont {D.~B.}\ \bibnamefont {Leinweber}}, \bibinfo
		{author} {\bibfnamefont {A.~W.}\ \bibnamefont {Thomas}},\ and\ \bibinfo
		{author} {\bibfnamefont {R.~D.}\ \bibnamefont {Young}},\ }\bibfield  {title}
	{\bibinfo {title} {Chiral and continuum extrapolation of partially-quenched
			lattice results},\ }\href
	{https://doi.org/https://doi.org/10.1016/j.physletb.2005.09.020} {\bibfield
		{journal} {\bibinfo  {journal} {Phys. Lett. B}\ }\textbf {\bibinfo {volume}
			{628}},\ \bibinfo {pages} {125} (\bibinfo {year} {2005})}\BibitemShut
	{NoStop}%
	\bibitem [{\citenamefont {Wang}\ \emph
		{et~al.}(2009{\natexlab{b}})\citenamefont {Wang}, \citenamefont {Leinweber},
		\citenamefont {Thomas},\ and\ \citenamefont {Young}}]{Wang}%
	\BibitemOpen
	\bibfield  {author} {\bibinfo {author} {\bibfnamefont {P.}~\bibnamefont
			{Wang}}, \bibinfo {author} {\bibfnamefont {D.}~\bibnamefont {Leinweber}},
		\bibinfo {author} {\bibfnamefont {A.}~\bibnamefont {Thomas}},\ and\ \bibinfo
		{author} {\bibfnamefont {R.}~\bibnamefont {Young}},\ }\bibfield  {title}
	{\bibinfo {title} {{Chiral extrapolation of octet-baryon charge radii}},\
	}\href {https://doi.org/10.1103/PhysRevD.79.094001} {\bibfield  {journal}
		{\bibinfo  {journal} {Phys. Rev. D}\ }\textbf {\bibinfo {volume} {79}},\
		\bibinfo {pages} {094001} (\bibinfo {year} {2009}{\natexlab{b}})},\ \Eprint
	{https://arxiv.org/abs/0810.1021} {arXiv:0810.1021 [hep-ph]} \BibitemShut
	{NoStop}%
	\bibitem [{\citenamefont {He}\ and\ \citenamefont
		{Wang}(2018{\natexlab{a}})}]{fw.nucleon}%
	\BibitemOpen
	\bibfield  {author} {\bibinfo {author} {\bibfnamefont {F.}~\bibnamefont
			{He}}\ and\ \bibinfo {author} {\bibfnamefont {P.}~\bibnamefont {Wang}},\
	}\bibfield  {title} {\bibinfo {title} {Nucleon electromagnetic form factors
			with a nonlocal chiral effective lagrangian},\ }\href
	{https://doi.org/10.1103/PhysRevD.97.036007} {\bibfield  {journal} {\bibinfo
			{journal} {Phys. Rev. D}\ }\textbf {\bibinfo {volume} {97}},\ \bibinfo
		{pages} {036007} (\bibinfo {year} {2018}{\natexlab{a}})}\BibitemShut
	{NoStop}%
	\bibitem [{\citenamefont {He}\ and\ \citenamefont
		{Wang}(2018{\natexlab{b}})}]{fw.strange}%
	\BibitemOpen
	\bibfield  {author} {\bibinfo {author} {\bibfnamefont {F.}~\bibnamefont
			{He}}\ and\ \bibinfo {author} {\bibfnamefont {P.}~\bibnamefont {Wang}},\
	}\bibfield  {title} {\bibinfo {title} {Strange form factors of the nucleon
			with a nonlocal chiral effective lagrangian},\ }\href {https://doi.org/ARTN
		036007 10.1103/PhysRevD.98.036007} {\bibfield  {journal} {\bibinfo  {journal}
			{Phys. Rev. D}\ }\textbf {\bibinfo {volume} {98}},\ \bibinfo {pages} {036007}
		(\bibinfo {year} {2018}{\natexlab{b}})}\BibitemShut {NoStop}%
	\bibitem [{\citenamefont {Salamu}\ \emph {et~al.}(2019)\citenamefont {Salamu},
		\citenamefont {Ji}, \citenamefont {Melnitchouk}, \citenamefont {Thomas},\
		and\ \citenamefont {Wang}}]{Salamu.parton}%
	\BibitemOpen
	\bibfield  {author} {\bibinfo {author} {\bibfnamefont {Y.}~\bibnamefont
			{Salamu}}, \bibinfo {author} {\bibfnamefont {C.-R.}\ \bibnamefont {Ji}},
		\bibinfo {author} {\bibfnamefont {W.}~\bibnamefont {Melnitchouk}}, \bibinfo
		{author} {\bibfnamefont {A.~W.}\ \bibnamefont {Thomas}},\ and\ \bibinfo
		{author} {\bibfnamefont {P.}~\bibnamefont {Wang}},\ }\bibfield  {title}
	{\bibinfo {title} {{Parton distributions from nonlocal chiral SU(3) effective
				theory: Splitting functions}},\ }\href {https://doi.org/ARTN 014041
		10.1103/PhysRevD.99.014041} {\bibfield  {journal} {\bibinfo  {journal} {Phys.
				Rev. D}\ }\textbf {\bibinfo {volume} {99}},\ \bibinfo {pages} {014041}
		(\bibinfo {year} {2019})}\BibitemShut {NoStop}%
	\bibitem [{\citenamefont {He}\ and\ \citenamefont {Wang}(2019)}]{He}%
	\BibitemOpen
	\bibfield  {author} {\bibinfo {author} {\bibfnamefont {F.}~\bibnamefont
			{He}}\ and\ \bibinfo {author} {\bibfnamefont {P.}~\bibnamefont {Wang}},\
	}\bibfield  {title} {\bibinfo {title} {{Sivers distribution functions of sea
				quark in proton with chiral Lagrangian}},\ }\href
	{https://doi.org/10.1103/PhysRevD.100.074032} {\bibfield  {journal} {\bibinfo
			{journal} {Phys. Rev. D}\ }\textbf {\bibinfo {volume} {100}},\ \bibinfo
		{pages} {074032} (\bibinfo {year} {2019})},\ \Eprint
	{https://arxiv.org/abs/1904.06815} {arXiv:1904.06815 [hep-ph]} \BibitemShut
	{NoStop}%
	\bibitem [{\citenamefont {He}\ and\ \citenamefont {Wang}(2020)}]{He2}%
	\BibitemOpen
	\bibfield  {author} {\bibinfo {author} {\bibfnamefont {F.}~\bibnamefont
			{He}}\ and\ \bibinfo {author} {\bibfnamefont {P.}~\bibnamefont {Wang}},\
	}\bibfield  {title} {\bibinfo {title} {{Pauli form factors of electron and
				muon in nonlocal quantum electrodynamics}},\ }\href
	{https://doi.org/10.1140/epjp/s13360-020-00151-y} {\bibfield  {journal}
		{\bibinfo  {journal} {Eur. Phys. J. Plus}\ }\textbf {\bibinfo {volume}
			{135}},\ \bibinfo {pages} {156} (\bibinfo {year} {2020})},\ \Eprint
	{https://arxiv.org/abs/1901.00271} {arXiv:1901.00271 [hep-ph]} \BibitemShut
	{NoStop}%
	\bibitem [{\citenamefont {Boinepalli}\ \emph {et~al.}(2006)\citenamefont
		{Boinepalli}, \citenamefont {Leinweber}, \citenamefont {Williams},
		\citenamefont {Zanotti},\ and\ \citenamefont {Zhang}}]{Boinepalli}%
	\BibitemOpen
	\bibfield  {author} {\bibinfo {author} {\bibfnamefont {S.}~\bibnamefont
			{Boinepalli}}, \bibinfo {author} {\bibfnamefont {D.}~\bibnamefont
			{Leinweber}}, \bibinfo {author} {\bibfnamefont {A.}~\bibnamefont {Williams}},
		\bibinfo {author} {\bibfnamefont {J.}~\bibnamefont {Zanotti}},\ and\ \bibinfo
		{author} {\bibfnamefont {J.}~\bibnamefont {Zhang}},\ }\bibfield  {title}
	{\bibinfo {title} {{Precision electromagnetic structure of octet baryons in
				the chiral regime}},\ }\href {https://doi.org/10.1103/PhysRevD.74.093005}
	{\bibfield  {journal} {\bibinfo  {journal} {Phys. Rev. D}\ }\textbf {\bibinfo
			{volume} {74}},\ \bibinfo {pages} {093005} (\bibinfo {year} {2006})},\
	\Eprint {https://arxiv.org/abs/hep-lat/0604022} {arXiv:hep-lat/0604022}
	\BibitemShut {NoStop}%
	\bibitem [{\citenamefont {Shanahan}\ \emph
		{et~al.}(2014{\natexlab{a}})\citenamefont {Shanahan}, \citenamefont {Thomas},
		\citenamefont {Young}, \citenamefont {Zanotti}, \citenamefont {Horsley},
		\citenamefont {Nakamura}, \citenamefont {Pleiter}, \citenamefont {Rakow},
		\citenamefont {Schierholz},\ and\ \citenamefont {St{\"u}ben}}]{Shanahan}%
	\BibitemOpen
	\bibfield  {author} {\bibinfo {author} {\bibfnamefont {P.}~\bibnamefont
			{Shanahan}}, \bibinfo {author} {\bibfnamefont {A.}~\bibnamefont {Thomas}},
		\bibinfo {author} {\bibfnamefont {R.}~\bibnamefont {Young}}, \bibinfo
		{author} {\bibfnamefont {J.}~\bibnamefont {Zanotti}}, \bibinfo {author}
		{\bibfnamefont {R.}~\bibnamefont {Horsley}}, \bibinfo {author} {\bibfnamefont
			{Y.}~\bibnamefont {Nakamura}}, \bibinfo {author} {\bibfnamefont
			{D.}~\bibnamefont {Pleiter}}, \bibinfo {author} {\bibfnamefont
			{P.}~\bibnamefont {Rakow}}, \bibinfo {author} {\bibfnamefont
			{G.}~\bibnamefont {Schierholz}},\ and\ \bibinfo {author} {\bibfnamefont
			{H.}~\bibnamefont {St{\"u}ben}} (\bibinfo {collaboration} {CSSM,
			QCDSF/UKQCD}),\ }\bibfield  {title} {\bibinfo {title} {{Magnetic form factors
				of the octet baryons from lattice QCD and chiral extrapolation}},\ }\href
	{https://doi.org/10.1103/PhysRevD.89.074511} {\bibfield  {journal} {\bibinfo
			{journal} {Phys. Rev. D}\ }\textbf {\bibinfo {volume} {89}},\ \bibinfo
		{pages} {074511} (\bibinfo {year} {2014}{\natexlab{a}})},\ \Eprint
	{https://arxiv.org/abs/1401.5862} {arXiv:1401.5862 [hep-lat]} \BibitemShut
	{NoStop}%
	\bibitem [{\citenamefont {Shanahan}\ \emph
		{et~al.}(2014{\natexlab{b}})\citenamefont {Shanahan}, \citenamefont {Thomas},
		\citenamefont {Young}, \citenamefont {Zanotti}, \citenamefont {Horsley},
		\citenamefont {Nakamura}, \citenamefont {Pleiter}, \citenamefont {Rakow},
		\citenamefont {Schierholz},\ and\ \citenamefont {St{\"u}ben}}]{Shanahan2}%
	\BibitemOpen
	\bibfield  {author} {\bibinfo {author} {\bibfnamefont {P.}~\bibnamefont
			{Shanahan}}, \bibinfo {author} {\bibfnamefont {A.}~\bibnamefont {Thomas}},
		\bibinfo {author} {\bibfnamefont {R.}~\bibnamefont {Young}}, \bibinfo
		{author} {\bibfnamefont {J.}~\bibnamefont {Zanotti}}, \bibinfo {author}
		{\bibfnamefont {R.}~\bibnamefont {Horsley}}, \bibinfo {author} {\bibfnamefont
			{Y.}~\bibnamefont {Nakamura}}, \bibinfo {author} {\bibfnamefont
			{D.}~\bibnamefont {Pleiter}}, \bibinfo {author} {\bibfnamefont
			{P.}~\bibnamefont {Rakow}}, \bibinfo {author} {\bibfnamefont
			{G.}~\bibnamefont {Schierholz}},\ and\ \bibinfo {author} {\bibfnamefont
			{H.}~\bibnamefont {St{\"u}ben}},\ }\bibfield  {title} {\bibinfo {title}
		{{Electric form factors of the octet baryons from lattice QCD and chiral
				extrapolation}},\ }\href {https://doi.org/10.1103/PhysRevD.90.034502}
	{\bibfield  {journal} {\bibinfo  {journal} {Phys. Rev. D}\ }\textbf {\bibinfo
			{volume} {90}},\ \bibinfo {pages} {034502} (\bibinfo {year}
		{2014}{\natexlab{b}})},\ \Eprint {https://arxiv.org/abs/1403.1965}
	{arXiv:1403.1965 [hep-lat]} \BibitemShut {NoStop}%
	\bibitem [{\citenamefont {Kubis}\ and\ \citenamefont
		{Meissner}(2001{\natexlab{b}})}]{Kubis}%
	\BibitemOpen
	\bibfield  {author} {\bibinfo {author} {\bibfnamefont {B.}~\bibnamefont
			{Kubis}}\ and\ \bibinfo {author} {\bibfnamefont {U.~G.}\ \bibnamefont
			{Meissner}},\ }\bibfield  {title} {\bibinfo {title} {{Baryon form-factors in
				chiral perturbation theory}},\ }\href {https://doi.org/10.1007/s100520100570}
	{\bibfield  {journal} {\bibinfo  {journal} {Eur. Phys. J. C}\ }\textbf
		{\bibinfo {volume} {18}},\ \bibinfo {pages} {747} (\bibinfo {year}
		{2001}{\natexlab{b}})},\ \Eprint {https://arxiv.org/abs/hep-ph/0010283}
	{arXiv:hep-ph/0010283} \BibitemShut {NoStop}%
	\bibitem [{\citenamefont {Xiao}\ \emph {et~al.}(2018)\citenamefont {Xiao},
		\citenamefont {Ren}, \citenamefont {Lu}, \citenamefont {Geng},\ and\
		\citenamefont {Meißner}}]{Xiao}%
	\BibitemOpen
	\bibfield  {author} {\bibinfo {author} {\bibfnamefont {Y.}~\bibnamefont
			{Xiao}}, \bibinfo {author} {\bibfnamefont {X.-L.}\ \bibnamefont {Ren}},
		\bibinfo {author} {\bibfnamefont {J.-X.}\ \bibnamefont {Lu}}, \bibinfo
		{author} {\bibfnamefont {L.-S.}\ \bibnamefont {Geng}},\ and\ \bibinfo
		{author} {\bibfnamefont {U.-G.}\ \bibnamefont {Meißner}},\ }\bibfield
	{title} {\bibinfo {title} {{Octet baryon magnetic moments at
				next-to-next-to-leading order in covariant chiral perturbation theory}},\
	}\href {https://doi.org/10.1140/epjc/s10052-018-5960-4} {\bibfield  {journal}
		{\bibinfo  {journal} {Eur. Phys. J. C}\ }\textbf {\bibinfo {volume} {78}},\
		\bibinfo {pages} {489} (\bibinfo {year} {2018})},\ \Eprint
	{https://arxiv.org/abs/1803.04251} {arXiv:1803.04251 [hep-ph]} \BibitemShut
	{NoStop}%
	\bibitem [{\citenamefont {Geng}\ \emph {et~al.}(2009)\citenamefont {Geng},
		\citenamefont {Martin~Camalich},\ and\ \citenamefont {Vicente~Vacas}}]{Geng}%
	\BibitemOpen
	\bibfield  {author} {\bibinfo {author} {\bibfnamefont {L.}~\bibnamefont
			{Geng}}, \bibinfo {author} {\bibfnamefont {J.}~\bibnamefont
			{Martin~Camalich}},\ and\ \bibinfo {author} {\bibfnamefont {M.}~\bibnamefont
			{Vicente~Vacas}},\ }\bibfield  {title} {\bibinfo {title} {{Leading-order
				decuplet contributions to the baryon magnetic moments in Chiral Perturbation
				Theory}},\ }\href {https://doi.org/10.1016/j.physletb.2009.04.061} {\bibfield
		{journal} {\bibinfo  {journal} {Phys. Lett. B}\ }\textbf {\bibinfo {volume}
			{676}},\ \bibinfo {pages} {63} (\bibinfo {year} {2009})},\ \Eprint
	{https://arxiv.org/abs/0903.0779} {arXiv:0903.0779 [hep-ph]} \BibitemShut
	{NoStop}%
	\bibitem [{\citenamefont {Hiller~Blin}(2017)}]{Blin}%
	\BibitemOpen
	\bibfield  {author} {\bibinfo {author} {\bibfnamefont {A.}~\bibnamefont
			{Hiller~Blin}},\ }\bibfield  {title} {\bibinfo {title} {{Systematic study of
				octet-baryon electromagnetic form factors in covariant chiral perturbation
				theory}},\ }\href {https://doi.org/10.1103/PhysRevD.96.093008} {\bibfield
		{journal} {\bibinfo  {journal} {Phys. Rev. D}\ }\textbf {\bibinfo {volume}
			{96}},\ \bibinfo {pages} {093008} (\bibinfo {year} {2017})},\ \Eprint
	{https://arxiv.org/abs/1707.02255} {arXiv:1707.02255 [hep-ph]} \BibitemShut
	{NoStop}%
	\bibitem [{\citenamefont {Jenkins}(1992)}]{Jenkins1}%
	\BibitemOpen
	\bibfield  {author} {\bibinfo {author} {\bibfnamefont {E.~E.}\ \bibnamefont
			{Jenkins}},\ }\bibfield  {title} {\bibinfo {title} {{Baryon masses in chiral
				perturbation theory}},\ }\href {https://doi.org/10.1016/0550-3213(92)90203-N}
	{\bibfield  {journal} {\bibinfo  {journal} {Nucl. Phys. B}\ }\textbf
		{\bibinfo {volume} {368}},\ \bibinfo {pages} {190} (\bibinfo {year}
		{1992})}\BibitemShut {NoStop}%
	\bibitem [{\citenamefont {Jenkins}\ \emph {et~al.}(1993)\citenamefont
		{Jenkins}, \citenamefont {Luke}, \citenamefont {Manohar},\ and\ \citenamefont
		{Savage}}]{Jenkins2}%
	\BibitemOpen
	\bibfield  {author} {\bibinfo {author} {\bibfnamefont {E.~E.}\ \bibnamefont
			{Jenkins}}, \bibinfo {author} {\bibfnamefont {M.~E.}\ \bibnamefont {Luke}},
		\bibinfo {author} {\bibfnamefont {A.~V.}\ \bibnamefont {Manohar}},\ and\
		\bibinfo {author} {\bibfnamefont {M.~J.}\ \bibnamefont {Savage}},\ }\bibfield
	{title} {\bibinfo {title} {Chiral perturbation theory analysis of the baryon
			magnetic moments},\ }\href {https://doi.org/10.1016/0370-2693(93)90430-P}
	{\bibfield  {journal} {\bibinfo  {journal} {Phys. Lett. B}\ }\textbf
		{\bibinfo {volume} {302}},\ \bibinfo {pages} {482} (\bibinfo {year}
		{1993})},\ \Eprint {https://arxiv.org/abs/hep-ph/9212226}
	{arXiv:hep-ph/9212226} \BibitemShut {NoStop}%
	\bibitem [{\citenamefont {Wang}(2014)}]{pingw.quantization}%
	\BibitemOpen
	\bibfield  {author} {\bibinfo {author} {\bibfnamefont {P.}~\bibnamefont
			{Wang}},\ }\bibfield  {title} {\bibinfo {title} {Solid quantization for
			nonpoint particles},\ }\href {https://doi.org/10.1139/cjp-2012-0395}
	{\bibfield  {journal} {\bibinfo  {journal} {Can. J. Phys.}\ }\textbf
		{\bibinfo {volume} {92}},\ \bibinfo {pages} {25} (\bibinfo {year}
		{2014})}\BibitemShut {NoStop}%
	\bibitem [{\citenamefont {Lin}\ and\ \citenamefont {Orginos}(2009)}]{Lin}%
	\BibitemOpen
	\bibfield  {author} {\bibinfo {author} {\bibfnamefont {H.-W.}\ \bibnamefont
			{Lin}}\ and\ \bibinfo {author} {\bibfnamefont {K.}~\bibnamefont {Orginos}},\
	}\bibfield  {title} {\bibinfo {title} {{Strange Baryon Electromagnetic Form
				Factors and SU(3) Flavor Symmetry Breaking}},\ }\href
	{https://doi.org/10.1103/PhysRevD.79.074507} {\bibfield  {journal} {\bibinfo
			{journal} {Phys. Rev. D}\ }\textbf {\bibinfo {volume} {79}},\ \bibinfo
		{pages} {074507} (\bibinfo {year} {2009})},\ \Eprint
	{https://arxiv.org/abs/0812.4456} {arXiv:0812.4456 [hep-lat]} \BibitemShut
	{NoStop}%
	\bibitem [{\citenamefont {Carrillo-Serrano}\ \emph {et~al.}(2016)\citenamefont
		{Carrillo-Serrano}, \citenamefont {Bentz}, \citenamefont {Clo{\"e}t},\ and\
		\citenamefont {Thomas}}]{Serrano}%
	\BibitemOpen
	\bibfield  {author} {\bibinfo {author} {\bibfnamefont {M.~E.}\ \bibnamefont
			{Carrillo-Serrano}}, \bibinfo {author} {\bibfnamefont {W.}~\bibnamefont
			{Bentz}}, \bibinfo {author} {\bibfnamefont {I.~C.}\ \bibnamefont
			{Clo{\"e}t}},\ and\ \bibinfo {author} {\bibfnamefont {A.~W.}\ \bibnamefont
			{Thomas}},\ }\bibfield  {title} {\bibinfo {title} {{Baryon Octet
				Electromagnetic Form Factors in a confining NJL model}},\ }\href
	{https://doi.org/10.1016/j.physletb.2016.05.065} {\bibfield  {journal}
		{\bibinfo  {journal} {Phys. Lett. B}\ }\textbf {\bibinfo {volume} {759}},\
		\bibinfo {pages} {178} (\bibinfo {year} {2016})},\ \Eprint
	{https://arxiv.org/abs/1603.02741} {arXiv:1603.02741 [nucl-th]} \BibitemShut
	{NoStop}%
	\bibitem [{\citenamefont {Liu}\ \emph {et~al.}(2014)\citenamefont {Liu},
		\citenamefont {Khosonthongkee}, \citenamefont {Limphirat},\ and\
		\citenamefont {Yan}}]{Liu}%
	\BibitemOpen
	\bibfield  {author} {\bibinfo {author} {\bibfnamefont {X.}~\bibnamefont
			{Liu}}, \bibinfo {author} {\bibfnamefont {K.}~\bibnamefont {Khosonthongkee}},
		\bibinfo {author} {\bibfnamefont {A.}~\bibnamefont {Limphirat}},\ and\
		\bibinfo {author} {\bibfnamefont {Y.}~\bibnamefont {Yan}},\ }\bibfield
	{title} {\bibinfo {title} {{Study of baryon octet electromagnetic form
				factors in perturbative chiral quark model}},\ }\href
	{https://doi.org/10.1088/0954-3899/41/5/055008} {\bibfield  {journal}
		{\bibinfo  {journal} {J. Phys. G}\ }\textbf {\bibinfo {volume} {41}},\
		\bibinfo {pages} {055008} (\bibinfo {year} {2014})},\ \Eprint
	{https://arxiv.org/abs/1309.2063} {arXiv:1309.2063 [hep-ph]} \BibitemShut
	{NoStop}%
	\bibitem [{\citenamefont {Tanabashi}\ \emph {et~al.}(2018)\citenamefont
		{Tanabashi} \emph {et~al.}}]{PDG}%
	\BibitemOpen
	\bibfield  {author} {\bibinfo {author} {\bibfnamefont {M.}~\bibnamefont
			{Tanabashi}} \emph {et~al.} (\bibinfo {collaboration} {Particle Data
			Group}),\ }\bibfield  {title} {\bibinfo {title} {{Review of Particle
				Physics}},\ }\href {https://doi.org/10.1103/PhysRevD.98.030001} {\bibfield
		{journal} {\bibinfo  {journal} {Phys. Rev. D}\ }\textbf {\bibinfo {volume}
			{98}},\ \bibinfo {pages} {030001} (\bibinfo {year} {2018})}\BibitemShut
	{NoStop}%
	\bibitem [{\citenamefont {Borasoy}\ and\ \citenamefont
		{Meissner}(1997)}]{Borasoy}%
	\BibitemOpen
	\bibfield  {author} {\bibinfo {author} {\bibfnamefont {B.}~\bibnamefont
			{Borasoy}}\ and\ \bibinfo {author} {\bibfnamefont {U.-G.}\ \bibnamefont
			{Meissner}},\ }\bibfield  {title} {\bibinfo {title} {{Chiral Expansion of
				Baryon Masses and $\sigma$-Terms}},\ }\href
	{https://doi.org/10.1006/aphy.1996.5630} {\bibfield  {journal} {\bibinfo
			{journal} {Annals Phys.}\ }\textbf {\bibinfo {volume} {254}},\ \bibinfo
		{pages} {192} (\bibinfo {year} {1997})},\ \Eprint
	{https://arxiv.org/abs/hep-ph/9607432} {arXiv:hep-ph/9607432} \BibitemShut
	{NoStop}%
	\bibitem [{\citenamefont {Luty}\ and\ \citenamefont {White}(1993)}]{Luty}%
	\BibitemOpen
	\bibfield  {author} {\bibinfo {author} {\bibfnamefont {M.}~\bibnamefont
			{Luty}}\ and\ \bibinfo {author} {\bibfnamefont {M.~J.}\ \bibnamefont
			{White}},\ }\bibfield  {title} {\bibinfo {title} {{Decouplet contributions to
				hyperon axial vector form-factors}},\ }\href
	{https://doi.org/10.1016/0370-2693(93)90812-V} {\bibfield  {journal}
		{\bibinfo  {journal} {Phys. Lett. B}\ }\textbf {\bibinfo {volume} {319}},\
		\bibinfo {pages} {261} (\bibinfo {year} {1993})},\ \Eprint
	{https://arxiv.org/abs/hep-ph/9305203} {arXiv:hep-ph/9305203} \BibitemShut
	{NoStop}%
	\bibitem [{\citenamefont {Nath}\ \emph {et~al.}(1971)\citenamefont {Nath},
		\citenamefont {Etemadi},\ and\ \citenamefont {Kimel}}]{nath.decuplet}%
	\BibitemOpen
	\bibfield  {author} {\bibinfo {author} {\bibfnamefont {L.~M.}\ \bibnamefont
			{Nath}}, \bibinfo {author} {\bibfnamefont {B.}~\bibnamefont {Etemadi}},\ and\
		\bibinfo {author} {\bibfnamefont {J.~D.}\ \bibnamefont {Kimel}},\ }\bibfield
	{title} {\bibinfo {title} {Uniqueness of the interaction involving
			spin-32particles},\ }\href {https://doi.org/10.1103/PhysRevD.3.2153}
	{\bibfield  {journal} {\bibinfo  {journal} {Phys. Rev. D}\ }\textbf {\bibinfo
			{volume} {3}},\ \bibinfo {pages} {2153} (\bibinfo {year} {1971})}\BibitemShut
	{NoStop}%
	\bibitem [{\citenamefont {Janssens}\ \emph {et~al.}(1966)\citenamefont
		{Janssens}, \citenamefont {Hofstadter}, \citenamefont {Hughes},\ and\
		\citenamefont {Yearian}}]{Janssens1966}%
	\BibitemOpen
	\bibfield  {author} {\bibinfo {author} {\bibfnamefont {T.}~\bibnamefont
			{Janssens}}, \bibinfo {author} {\bibfnamefont {R.}~\bibnamefont
			{Hofstadter}}, \bibinfo {author} {\bibfnamefont {E.}~\bibnamefont {Hughes}},\
		and\ \bibinfo {author} {\bibfnamefont {M.}~\bibnamefont {Yearian}},\
	}\bibfield  {title} {\bibinfo {title} {{Proton form factors from elastic
				electron-proton scattering}},\ }\href
	{https://doi.org/10.1103/PhysRev.142.922} {\bibfield  {journal} {\bibinfo
			{journal} {Phys. Rev.}\ }\textbf {\bibinfo {volume} {142}},\ \bibinfo {pages}
		{922} (\bibinfo {year} {1966})}\BibitemShut {NoStop}%
	\bibitem [{\citenamefont {Berger}\ \emph {et~al.}(1971)\citenamefont {Berger},
		\citenamefont {Burkert}, \citenamefont {Knop}, \citenamefont {Langenbeck},\
		and\ \citenamefont {Rith}}]{Berger1971}%
	\BibitemOpen
	\bibfield  {author} {\bibinfo {author} {\bibfnamefont {C.}~\bibnamefont
			{Berger}}, \bibinfo {author} {\bibfnamefont {V.}~\bibnamefont {Burkert}},
		\bibinfo {author} {\bibfnamefont {G.}~\bibnamefont {Knop}}, \bibinfo {author}
		{\bibfnamefont {B.}~\bibnamefont {Langenbeck}},\ and\ \bibinfo {author}
		{\bibfnamefont {K.}~\bibnamefont {Rith}},\ }\bibfield  {title} {\bibinfo
		{title} {{Electromagnetic form-factors of the proton at squared four momentum
				transfers between 10-fm**-2 and 50 fm**-2}},\ }\href
	{https://doi.org/10.1016/0370-2693(71)90448-5} {\bibfield  {journal}
		{\bibinfo  {journal} {Phys. Lett. B}\ }\textbf {\bibinfo {volume} {35}},\
		\bibinfo {pages} {87} (\bibinfo {year} {1971})}\BibitemShut {NoStop}%
	\bibitem [{\citenamefont {Price}\ \emph {et~al.}(1971)\citenamefont {Price},
		\citenamefont {Dunning}, \citenamefont {Goitein}, \citenamefont {Hanson},
		\citenamefont {Kirk},\ and\ \citenamefont {Wilson}}]{Price1971}%
	\BibitemOpen
	\bibfield  {author} {\bibinfo {author} {\bibfnamefont {L.}~\bibnamefont
			{Price}}, \bibinfo {author} {\bibfnamefont {J.}~\bibnamefont {Dunning}},
		\bibinfo {author} {\bibfnamefont {M.}~\bibnamefont {Goitein}}, \bibinfo
		{author} {\bibfnamefont {K.}~\bibnamefont {Hanson}}, \bibinfo {author}
		{\bibfnamefont {T.}~\bibnamefont {Kirk}},\ and\ \bibinfo {author}
		{\bibfnamefont {R.}~\bibnamefont {Wilson}},\ }\bibfield  {title} {\bibinfo
		{title} {{Backward-angle electron-proton elastic scattering and proton
				electromagnetic form-factors}},\ }\href
	{https://doi.org/10.1103/PhysRevD.4.45} {\bibfield  {journal} {\bibinfo
			{journal} {Phys. Rev. D}\ }\textbf {\bibinfo {volume} {4}},\ \bibinfo {pages}
		{45} (\bibinfo {year} {1971})}\BibitemShut {NoStop}%
	\bibitem [{\citenamefont {Anklin}\ \emph {et~al.}(1994)\citenamefont {Anklin}
		\emph {et~al.}}]{Anklin1994}%
	\BibitemOpen
	\bibfield  {author} {\bibinfo {author} {\bibfnamefont {H.}~\bibnamefont
			{Anklin}} \emph {et~al.},\ }\bibfield  {title} {\bibinfo {title} {{Precision
				measurement of the neutron magnetic form-factor}},\ }\href
	{https://doi.org/10.1016/0370-2693(94)90538-X} {\bibfield  {journal}
		{\bibinfo  {journal} {Phys. Lett. B}\ }\textbf {\bibinfo {volume} {336}},\
		\bibinfo {pages} {313} (\bibinfo {year} {1994})}\BibitemShut {NoStop}%
	\bibitem [{\citenamefont {Walker}\ \emph {et~al.}(1994)\citenamefont {Walker}
		\emph {et~al.}}]{Walker1994}%
	\BibitemOpen
	\bibfield  {author} {\bibinfo {author} {\bibfnamefont {R.}~\bibnamefont
			{Walker}} \emph {et~al.},\ }\bibfield  {title} {\bibinfo {title}
		{{Measurements of the proton elastic form-factors for 1-GeV/c**2 <= Q**2 <=
				3-GeV/C**2 at SLAC}},\ }\href {https://doi.org/10.1103/PhysRevD.49.5671}
	{\bibfield  {journal} {\bibinfo  {journal} {Phys. Rev. D}\ }\textbf {\bibinfo
			{volume} {49}},\ \bibinfo {pages} {5671} (\bibinfo {year}
		{1994})}\BibitemShut {NoStop}%
	\bibitem [{\citenamefont {Bartel}\ \emph {et~al.}(1973)\citenamefont {Bartel},
		\citenamefont {Busser}, \citenamefont {Dix}, \citenamefont {Felst},
		\citenamefont {Harms}, \citenamefont {Krehbiel}, \citenamefont {Kuhlmann},
		\citenamefont {McElroy}, \citenamefont {Meyer},\ and\ \citenamefont
		{Weber}}]{Bartel1973}%
	\BibitemOpen
	\bibfield  {author} {\bibinfo {author} {\bibfnamefont {W.}~\bibnamefont
			{Bartel}}, \bibinfo {author} {\bibfnamefont {F.}~\bibnamefont {Busser}},
		\bibinfo {author} {\bibfnamefont {W.}~\bibnamefont {Dix}}, \bibinfo {author}
		{\bibfnamefont {R.}~\bibnamefont {Felst}}, \bibinfo {author} {\bibfnamefont
			{D.}~\bibnamefont {Harms}}, \bibinfo {author} {\bibfnamefont
			{H.}~\bibnamefont {Krehbiel}}, \bibinfo {author} {\bibfnamefont
			{P.}~\bibnamefont {Kuhlmann}}, \bibinfo {author} {\bibfnamefont
			{J.}~\bibnamefont {McElroy}}, \bibinfo {author} {\bibfnamefont
			{J.}~\bibnamefont {Meyer}},\ and\ \bibinfo {author} {\bibfnamefont
			{G.}~\bibnamefont {Weber}},\ }\bibfield  {title} {\bibinfo {title}
		{{Measurement of proton and neutron electromagnetic form-factors at squared
				four momentum transfers up to 3-GeV/c$^2$}},\ }\href
	{https://doi.org/10.1016/0550-3213(73)90594-4} {\bibfield  {journal}
		{\bibinfo  {journal} {Nucl. Phys. B}\ }\textbf {\bibinfo {volume} {58}},\
		\bibinfo {pages} {429} (\bibinfo {year} {1973})}\BibitemShut {NoStop}%
	\bibitem [{\citenamefont {Arrington}\ \emph {et~al.}(2007)\citenamefont
		{Arrington}, \citenamefont {Melnitchouk},\ and\ \citenamefont
		{Tjon}}]{Arrington2007}%
	\BibitemOpen
	\bibfield  {author} {\bibinfo {author} {\bibfnamefont {J.}~\bibnamefont
			{Arrington}}, \bibinfo {author} {\bibfnamefont {W.}~\bibnamefont
			{Melnitchouk}},\ and\ \bibinfo {author} {\bibfnamefont {J.~A.}\ \bibnamefont
			{Tjon}},\ }\bibfield  {title} {\bibinfo {title} {Global analysis of proton
			elastic form factor data with two-photon exchange corrections},\ }\href
	{https://doi.org/10.1103/PhysRevC.76.035205} {\bibfield  {journal} {\bibinfo
			{journal} {Phys. Rev. C}\ }\textbf {\bibinfo {volume} {76}},\ \bibinfo
		{pages} {035205} (\bibinfo {year} {2007})}\BibitemShut {NoStop}%
	\bibitem [{\citenamefont {Golak}\ \emph {et~al.}(2001)\citenamefont {Golak},
		\citenamefont {Ziemer}, \citenamefont {Kamada}, \citenamefont {Witala},\ and\
		\citenamefont {Gloeckle}}]{Golak2001}%
	\BibitemOpen
	\bibfield  {author} {\bibinfo {author} {\bibfnamefont {J.}~\bibnamefont
			{Golak}}, \bibinfo {author} {\bibfnamefont {G.}~\bibnamefont {Ziemer}},
		\bibinfo {author} {\bibfnamefont {H.}~\bibnamefont {Kamada}}, \bibinfo
		{author} {\bibfnamefont {H.}~\bibnamefont {Witala}},\ and\ \bibinfo {author}
		{\bibfnamefont {W.}~\bibnamefont {Gloeckle}},\ }\bibfield  {title} {\bibinfo
		{title} {{Extraction of electromagnetic neutron form-factors through
				inclusive and exclusive polarized electron scattering on polarized He-3
				target}},\ }\href {https://doi.org/10.1103/PhysRevC.63.034006} {\bibfield
		{journal} {\bibinfo  {journal} {Phys. Rev. C}\ }\textbf {\bibinfo {volume}
			{63}},\ \bibinfo {pages} {034006} (\bibinfo {year} {2001})},\ \Eprint
	{https://arxiv.org/abs/nucl-th/0008008} {arXiv:nucl-th/0008008} \BibitemShut
	{NoStop}%
	\bibitem [{\citenamefont {Markowitz}\ \emph {et~al.}(1993)\citenamefont
		{Markowitz} \emph {et~al.}}]{Markowitz1993}%
	\BibitemOpen
	\bibfield  {author} {\bibinfo {author} {\bibfnamefont {P.}~\bibnamefont
			{Markowitz}} \emph {et~al.},\ }\bibfield  {title} {\bibinfo {title}
		{{Measurement of the magnetic form factor of the neutron}},\ }\href
	{https://doi.org/10.1103/PhysRevC.48.R5} {\bibfield  {journal} {\bibinfo
			{journal} {Phys. Rev. C}\ }\textbf {\bibinfo {volume} {48}},\ \bibinfo
		{pages} {5} (\bibinfo {year} {1993})}\BibitemShut {NoStop}%
	\bibitem [{\citenamefont {Bruins}\ \emph {et~al.}(1995)\citenamefont {Bruins}
		\emph {et~al.}}]{Bruins1995}%
	\BibitemOpen
	\bibfield  {author} {\bibinfo {author} {\bibfnamefont {E.}~\bibnamefont
			{Bruins}} \emph {et~al.},\ }\bibfield  {title} {\bibinfo {title}
		{{Measurement of the neutron magnetic form-factor}},\ }\href
	{https://doi.org/10.1103/PhysRevLett.75.21} {\bibfield  {journal} {\bibinfo
			{journal} {Phys. Rev. Lett.}\ }\textbf {\bibinfo {volume} {75}},\ \bibinfo
		{pages} {21} (\bibinfo {year} {1995})}\BibitemShut {NoStop}%
	\bibitem [{\citenamefont {Anklin}\ \emph {et~al.}(1998)\citenamefont {Anklin}
		\emph {et~al.}}]{Anklin1998}%
	\BibitemOpen
	\bibfield  {author} {\bibinfo {author} {\bibfnamefont {H.}~\bibnamefont
			{Anklin}} \emph {et~al.},\ }\bibfield  {title} {\bibinfo {title} {{Precise
				measurements of the neutron magnetic form-factor}},\ }\href
	{https://doi.org/10.1016/S0370-2693(98)00442-0} {\bibfield  {journal}
		{\bibinfo  {journal} {Phys. Lett. B}\ }\textbf {\bibinfo {volume} {428}},\
		\bibinfo {pages} {248} (\bibinfo {year} {1998})}\BibitemShut {NoStop}%
	\bibitem [{\citenamefont {Xu}\ \emph {et~al.}(2000)\citenamefont {Xu} \emph
		{et~al.}}]{Xu2000}%
	\BibitemOpen
	\bibfield  {author} {\bibinfo {author} {\bibfnamefont {W.}~\bibnamefont {Xu}}
		\emph {et~al.},\ }\bibfield  {title} {\bibinfo {title} {{The Transverse
				asymmetry A(T-prime) from quasielastic polarized He-3 (polarized e, e-prime)
				process and the neutron magnetic form-factor}},\ }\href
	{https://doi.org/10.1103/PhysRevLett.85.2900} {\bibfield  {journal} {\bibinfo
			{journal} {Phys. Rev. Lett.}\ }\textbf {\bibinfo {volume} {85}},\ \bibinfo
		{pages} {2900} (\bibinfo {year} {2000})},\ \Eprint
	{https://arxiv.org/abs/nucl-ex/0008003} {arXiv:nucl-ex/0008003} \BibitemShut
	{NoStop}%
	\bibitem [{\citenamefont {Kubon}\ \emph {et~al.}(2002)\citenamefont {Kubon}
		\emph {et~al.}}]{Kubon2002}%
	\BibitemOpen
	\bibfield  {author} {\bibinfo {author} {\bibfnamefont {G.}~\bibnamefont
			{Kubon}} \emph {et~al.},\ }\bibfield  {title} {\bibinfo {title} {{Precise
				neutron magnetic form-factors}},\ }\href
	{https://doi.org/10.1016/S0370-2693(01)01386-7} {\bibfield  {journal}
		{\bibinfo  {journal} {Phys. Lett. B}\ }\textbf {\bibinfo {volume} {524}},\
		\bibinfo {pages} {26} (\bibinfo {year} {2002})},\ \Eprint
	{https://arxiv.org/abs/nucl-ex/0107016} {arXiv:nucl-ex/0107016} \BibitemShut
	{NoStop}%
	\bibitem [{\citenamefont {Madey}\ \emph {et~al.}(2003)\citenamefont {Madey}
		\emph {et~al.}}]{Madey2003}%
	\BibitemOpen
	\bibfield  {author} {\bibinfo {author} {\bibfnamefont {R.}~\bibnamefont
			{Madey}} \emph {et~al.} (\bibinfo {collaboration} {E93-038}),\ }\bibfield
	{title} {\bibinfo {title} {{Measurements of G(E)n / G(M)n from the
				H-2(polarized-e,e-prime polarized-n) reaction to Q**2 = 1.45 (GeV/c)**2}},\
	}\href {https://doi.org/10.1103/PhysRevLett.91.122002} {\bibfield  {journal}
		{\bibinfo  {journal} {Phys. Rev. Lett.}\ }\textbf {\bibinfo {volume} {91}},\
		\bibinfo {pages} {122002} (\bibinfo {year} {2003})},\ \Eprint
	{https://arxiv.org/abs/nucl-ex/0308007} {arXiv:nucl-ex/0308007} \BibitemShut
	{NoStop}%
	\bibitem [{\citenamefont {Xu}\ \emph {et~al.}(2003)\citenamefont {Xu} \emph
		{et~al.}}]{Xu2003}%
	\BibitemOpen
	\bibfield  {author} {\bibinfo {author} {\bibfnamefont {W.}~\bibnamefont {Xu}}
		\emph {et~al.} (\bibinfo {collaboration} {Jefferson Lab E95-001}),\
	}\bibfield  {title} {\bibinfo {title} {{PWIA extraction of the neutron
				magnetic form-factor from quasielastic polarized-He-3(polarized-e, e-prime)
				at Q**2 = 0.3-(GeV/c)**2 - 0.6-(GeV/c)**2}},\ }\href
	{https://doi.org/10.1103/PhysRevC.67.012201} {\bibfield  {journal} {\bibinfo
			{journal} {Phys. Rev. C}\ }\textbf {\bibinfo {volume} {67}},\ \bibinfo
		{pages} {012201} (\bibinfo {year} {2003})},\ \Eprint
	{https://arxiv.org/abs/nucl-ex/0208007} {arXiv:nucl-ex/0208007} \BibitemShut
	{NoStop}%
	\bibitem [{\citenamefont {Bauer}\ \emph {et~al.}(2012)\citenamefont {Bauer},
		\citenamefont {Bernauer},\ and\ \citenamefont {Scherer}}]{Bauer}%
	\BibitemOpen
	\bibfield  {author} {\bibinfo {author} {\bibfnamefont {T.}~\bibnamefont
			{Bauer}}, \bibinfo {author} {\bibfnamefont {J.}~\bibnamefont {Bernauer}},\
		and\ \bibinfo {author} {\bibfnamefont {S.}~\bibnamefont {Scherer}},\
	}\bibfield  {title} {\bibinfo {title} {{Electromagnetic form factors of the
				nucleon in effective field theory}},\ }\href
	{https://doi.org/10.1103/PhysRevC.86.065206} {\bibfield  {journal} {\bibinfo
			{journal} {Phys. Rev. C}\ }\textbf {\bibinfo {volume} {86}},\ \bibinfo
		{pages} {065206} (\bibinfo {year} {2012})},\ \Eprint
	{https://arxiv.org/abs/1209.3872} {arXiv:1209.3872 [nucl-th]} \BibitemShut
	{NoStop}%
	\bibitem [{\citenamefont {Hanson}\ \emph {et~al.}(1973)\citenamefont {Hanson},
		\citenamefont {Dunning}, \citenamefont {Goitein}, \citenamefont {Kirk},
		\citenamefont {Price},\ and\ \citenamefont {Wilson}}]{Hanson1973}%
	\BibitemOpen
	\bibfield  {author} {\bibinfo {author} {\bibfnamefont {K.}~\bibnamefont
			{Hanson}}, \bibinfo {author} {\bibfnamefont {J.}~\bibnamefont {Dunning}},
		\bibinfo {author} {\bibfnamefont {M.}~\bibnamefont {Goitein}}, \bibinfo
		{author} {\bibfnamefont {T.}~\bibnamefont {Kirk}}, \bibinfo {author}
		{\bibfnamefont {L.}~\bibnamefont {Price}},\ and\ \bibinfo {author}
		{\bibfnamefont {R.}~\bibnamefont {Wilson}},\ }\bibfield  {title} {\bibinfo
		{title} {{Large angle quasielastic electron-deuteron scattering}},\ }\href
	{https://doi.org/10.1103/PhysRevD.8.753} {\bibfield  {journal} {\bibinfo
			{journal} {Phys. Rev. D}\ }\textbf {\bibinfo {volume} {8}},\ \bibinfo {pages}
		{753} (\bibinfo {year} {1973})}\BibitemShut {NoStop}%
	\bibitem [{\citenamefont {Murphy}\ \emph {et~al.}(1974)\citenamefont {Murphy},
		\citenamefont {Shin},\ and\ \citenamefont {Skopik}}]{Murphy1974}%
	\BibitemOpen
	\bibfield  {author} {\bibinfo {author} {\bibfnamefont {J.}~\bibnamefont
			{Murphy}}, \bibinfo {author} {\bibfnamefont {Y.}~\bibnamefont {Shin}},\ and\
		\bibinfo {author} {\bibfnamefont {D.}~\bibnamefont {Skopik}},\ }\bibfield
	{title} {\bibinfo {title} {{Proton form factor from 0.15 to 0.79 fm-2}},\
	}\href {https://doi.org/10.1103/PhysRevC.9.2125} {\bibfield  {journal}
		{\bibinfo  {journal} {Phys. Rev. C}\ }\textbf {\bibinfo {volume} {9}},\
		\bibinfo {pages} {2125} (\bibinfo {year} {1974})},\ \bibinfo {note}
	{[Erratum: Phys.Rev.C 10, 2111--2111 (1974)]}\BibitemShut {NoStop}%
	\bibitem [{\citenamefont {Hohler}\ \emph {et~al.}(1976)\citenamefont {Hohler},
		\citenamefont {Pietarinen}, \citenamefont {Sabba~Stefanescu}, \citenamefont
		{Borkowski}, \citenamefont {Simon}, \citenamefont {Walther},\ and\
		\citenamefont {Wendling}}]{Hohler1976}%
	\BibitemOpen
	\bibfield  {author} {\bibinfo {author} {\bibfnamefont {G.}~\bibnamefont
			{Hohler}}, \bibinfo {author} {\bibfnamefont {E.}~\bibnamefont {Pietarinen}},
		\bibinfo {author} {\bibfnamefont {I.}~\bibnamefont {Sabba~Stefanescu}},
		\bibinfo {author} {\bibfnamefont {F.}~\bibnamefont {Borkowski}}, \bibinfo
		{author} {\bibfnamefont {G.}~\bibnamefont {Simon}}, \bibinfo {author}
		{\bibfnamefont {V.}~\bibnamefont {Walther}},\ and\ \bibinfo {author}
		{\bibfnamefont {R.}~\bibnamefont {Wendling}},\ }\bibfield  {title} {\bibinfo
		{title} {{Analysis of Electromagnetic Nucleon Form-Factors}},\ }\href
	{https://doi.org/10.1016/0550-3213(76)90449-1} {\bibfield  {journal}
		{\bibinfo  {journal} {Nucl. Phys. B}\ }\textbf {\bibinfo {volume} {114}},\
		\bibinfo {pages} {505} (\bibinfo {year} {1976})}\BibitemShut {NoStop}%
	\bibitem [{\citenamefont {Simon}\ \emph {et~al.}(1980)\citenamefont {Simon},
		\citenamefont {Schmitt}, \citenamefont {Borkowski},\ and\ \citenamefont
		{Walther}}]{Simon1980}%
	\BibitemOpen
	\bibfield  {author} {\bibinfo {author} {\bibfnamefont {G.}~\bibnamefont
			{Simon}}, \bibinfo {author} {\bibfnamefont {C.}~\bibnamefont {Schmitt}},
		\bibinfo {author} {\bibfnamefont {F.}~\bibnamefont {Borkowski}},\ and\
		\bibinfo {author} {\bibfnamefont {V.}~\bibnamefont {Walther}},\ }\bibfield
	{title} {\bibinfo {title} {{Absolute electron Proton Cross-Sections at Low
				Momentum Transfer Measured with a High Pressure Gas Target System}},\ }\href
	{https://doi.org/10.1016/0375-9474(80)90104-9} {\bibfield  {journal}
		{\bibinfo  {journal} {Nucl. Phys. A}\ }\textbf {\bibinfo {volume} {333}},\
		\bibinfo {pages} {381} (\bibinfo {year} {1980})}\BibitemShut {NoStop}%
	\bibitem [{\citenamefont {Eden}\ \emph {et~al.}(1994)\citenamefont {Eden} \emph
		{et~al.}}]{Eden1994}%
	\BibitemOpen
	\bibfield  {author} {\bibinfo {author} {\bibfnamefont {T.}~\bibnamefont
			{Eden}} \emph {et~al.},\ }\bibfield  {title} {\bibinfo {title} {{Electric
				form factor of the neutron from the ${}^{2}H(\vec{e},e' \vec{n})^{1}H$
				reaction at $Q^{2} = 0.255$ (GeV/c)${}^2$}},\ }\href
	{https://doi.org/10.1103/PhysRevC.50.R1749} {\bibfield  {journal} {\bibinfo
			{journal} {Phys. Rev. C}\ }\textbf {\bibinfo {volume} {50}},\ \bibinfo
		{pages} {1749} (\bibinfo {year} {1994})}\BibitemShut {NoStop}%
	\bibitem [{\citenamefont {Herberg}\ \emph {et~al.}(1999)\citenamefont {Herberg}
		\emph {et~al.}}]{Herberg1999}%
	\BibitemOpen
	\bibfield  {author} {\bibinfo {author} {\bibfnamefont {C.}~\bibnamefont
			{Herberg}} \emph {et~al.},\ }\bibfield  {title} {\bibinfo {title}
		{{Determination of the neutron electric form-factor in the D(e,e' n)p
				reaction and the influence of nuclear binding}},\ }\href
	{https://doi.org/10.1007/s100500050268} {\bibfield  {journal} {\bibinfo
			{journal} {Eur. Phys. J. A}\ }\textbf {\bibinfo {volume} {5}},\ \bibinfo
		{pages} {131} (\bibinfo {year} {1999})}\BibitemShut {NoStop}%
	\bibitem [{\citenamefont {Ostrick}\ \emph {et~al.}(1999)\citenamefont {Ostrick}
		\emph {et~al.}}]{Ostrick1999}%
	\BibitemOpen
	\bibfield  {author} {\bibinfo {author} {\bibfnamefont {M.}~\bibnamefont
			{Ostrick}} \emph {et~al.},\ }\bibfield  {title} {\bibinfo {title}
		{{Measurement of the neutron electric form-factor G(E,n) in the quasifree
				H-2(e(pol.),e' n(pol.))p reaction}},\ }\href
	{https://doi.org/10.1103/PhysRevLett.83.276} {\bibfield  {journal} {\bibinfo
			{journal} {Phys. Rev. Lett.}\ }\textbf {\bibinfo {volume} {83}},\ \bibinfo
		{pages} {276} (\bibinfo {year} {1999})}\BibitemShut {NoStop}%
	\bibitem [{\citenamefont {Passchier}\ \emph {et~al.}(1999)\citenamefont
		{Passchier} \emph {et~al.}}]{Passchier1999}%
	\BibitemOpen
	\bibfield  {author} {\bibinfo {author} {\bibfnamefont {I.}~\bibnamefont
			{Passchier}} \emph {et~al.},\ }\bibfield  {title} {\bibinfo {title} {{The
				Charge form-factor of the neutron from the reaction polarized H-2(polarized
				e, e-prime n) p}},\ }\href {https://doi.org/10.1103/PhysRevLett.82.4988}
	{\bibfield  {journal} {\bibinfo  {journal} {Phys. Rev. Lett.}\ }\textbf
		{\bibinfo {volume} {82}},\ \bibinfo {pages} {4988} (\bibinfo {year}
		{1999})},\ \Eprint {https://arxiv.org/abs/nucl-ex/9907012}
	{arXiv:nucl-ex/9907012} \BibitemShut {NoStop}%
	\bibitem [{\citenamefont {Bermuth}\ \emph {et~al.}(2003)\citenamefont {Bermuth}
		\emph {et~al.}}]{Bermuth2003}%
	\BibitemOpen
	\bibfield  {author} {\bibinfo {author} {\bibfnamefont {J.}~\bibnamefont
			{Bermuth}} \emph {et~al.},\ }\bibfield  {title} {\bibinfo {title} {{The
				Neutron charge form-factor and target analyzing powers from polarized-He-3
				(polarized-e,e-prime n) scattering}},\ }\href
	{https://doi.org/10.1016/S0370-2693(03)00725-1} {\bibfield  {journal}
		{\bibinfo  {journal} {Phys. Lett. B}\ }\textbf {\bibinfo {volume} {564}},\
		\bibinfo {pages} {199} (\bibinfo {year} {2003})},\ \Eprint
	{https://arxiv.org/abs/nucl-ex/0303015} {arXiv:nucl-ex/0303015} \BibitemShut
	{NoStop}%
	\bibitem [{\citenamefont {Warren}\ \emph {et~al.}(2004)\citenamefont {Warren}
		\emph {et~al.}}]{Warren2003}%
	\BibitemOpen
	\bibfield  {author} {\bibinfo {author} {\bibfnamefont {G.}~\bibnamefont
			{Warren}} \emph {et~al.} (\bibinfo {collaboration} {Jefferson Lab E93-026}),\
	}\bibfield  {title} {\bibinfo {title} {{Measurement of the electric
				form-factor of the neutron at $Q^2$ = 0.5 and 1.0 $GeV^2/c^2$}},\ }\href
	{https://doi.org/10.1103/PhysRevLett.92.042301} {\bibfield  {journal}
		{\bibinfo  {journal} {Phys. Rev. Lett.}\ }\textbf {\bibinfo {volume} {92}},\
		\bibinfo {pages} {042301} (\bibinfo {year} {2004})},\ \Eprint
	{https://arxiv.org/abs/nucl-ex/0308021} {arXiv:nucl-ex/0308021} \BibitemShut
	{NoStop}%
	\bibitem [{\citenamefont {Glazier}\ \emph {et~al.}(2005)\citenamefont {Glazier}
		\emph {et~al.}}]{Glazier2005}%
	\BibitemOpen
	\bibfield  {author} {\bibinfo {author} {\bibfnamefont {D.}~\bibnamefont
			{Glazier}} \emph {et~al.},\ }\bibfield  {title} {\bibinfo {title}
		{{Measurement of the electric form-factor of the neutron at Q**2 =
				0.3-(GeV/c)**2 to 0.8-(GeV/c)**2}},\ }\href
	{https://doi.org/10.1140/epja/i2004-10115-8} {\bibfield  {journal} {\bibinfo
			{journal} {Eur. Phys. J. A}\ }\textbf {\bibinfo {volume} {24}},\ \bibinfo
		{pages} {101} (\bibinfo {year} {2005})},\ \Eprint
	{https://arxiv.org/abs/nucl-ex/0410026} {arXiv:nucl-ex/0410026} \BibitemShut
	{NoStop}%
	\bibitem [{\citenamefont {Patel}(2015)}]{tool.packagex}%
	\BibitemOpen
	\bibfield  {author} {\bibinfo {author} {\bibfnamefont {H.~H.}\ \bibnamefont
			{Patel}},\ }\bibfield  {title} {\bibinfo {title} {{Package-X: A Mathematica
				package for the analytic calculation of one-loop integrals}},\ }\href
	{https://doi.org/10.1016/j.cpc.2015.08.017} {\bibfield  {journal} {\bibinfo
			{journal} {Comput. Phys. Commun.}\ }\textbf {\bibinfo {volume} {197}},\
		\bibinfo {pages} {276} (\bibinfo {year} {2015})}\BibitemShut {NoStop}%
\end{thebibliography}
%%%++++++++++++++++++

%

\end{document}